\documentclass[acmsmall,authorversion,screen]{acmart}
\usepackage{booktabs}
\usepackage{wrapfig}
\usepackage{todonotes}
\usepackage{algpseudocode}
\usepackage{float}
\usepackage{enumitem}
\usepackage{xspace}

\graphicspath{ {./figures/} }

\usepackage{titlesec}
\titlespacing*{\subsubsection}{\parindent}{1ex}{1em}

\usepackage[ruled,linesnumbered]{algorithm2e}

\SetAlFnt{\small}
\SetAlCapFnt{\small}
\SetAlCapNameFnt{\small}
\SetAlCapHSkip{0pt}

\SetCommentSty{mycommfont}

\acmJournal{TAAS}
\acmVolume{14}
\acmNumber{3}
\acmArticle{12}
\acmYear{2020}
\acmMonth{2}
\acmArticleSeq{1}
\acmPrice{15.00}
\acmDOI{10.1145/3380965}
\setcopyright{acmcopyright}

\newcommand{\mRUBiS}{\mbox{mRUBiS}\xspace}
\newcommand{\elem}[1]{\textsf{\small{#1}}}
\renewcommand{\elem}[1]{\textsf{{#1}}}
\newcommand{\scalefactor}{.86}
\newcommand{\Pa}[0]{\mathcal{P}}
\newcommand{\ra}[1]{\renewcommand{\arraystretch}{#1}}
\newcommand{\grid}{\emph{Grid5000}\xspace}
\newcommand{\deug}{\emph{DEUG}\xspace}
\newcommand{\lri}{\emph{LRI}\xspace}

\begin{document}
\title{Improving Scalability and Reward of Utility-Driven Self-Healing for Large Dynamic Architectures}

\author{Sona Ghahremani}
\email{sona.ghahremani@hpi.de}

\author{Holger Giese}
\affiliation{
	\institution{University of Potsdam}
}
\email{holger.giese@hpi.de}

\author{Thomas Vogel}
\orcid{0000-0002-7127-352X}
\affiliation{
	\institution{Humboldt-Universit\"at zu Berlin}
}
\email{thomas.vogel@cs.hu-berlin.de}

\begin{abstract}
Self-adaptation can be realized in various ways. Rule-based approaches prescribe the adaptation to be executed if the system or environment satisfies certain conditions. They result in scalable solutions but often with merely satisfying adaptation decisions. In contrast, utility-driven approaches determine optimal decisions by using an often costly optimization, which typically does not scale for large problems. We propose a rule-based and utility-driven adaptation scheme that achieves the benefits of both directions such that the adaptation decisions are optimal, whereas the computation scales by avoiding an expensive optimization. We use this adaptation scheme for architecture-based self-healing of large software systems. For this purpose, we define the utility for large dynamic architectures of such systems based on patterns that define issues the self-healing must address. Moreover, we use pattern-based adaptation rules to resolve these issues. Using a pattern-based scheme to define the utility and adaptation rules allows us to compute the impact of each rule application on the overall utility and to realize an incremental and efficient utility-driven self-healing. In addition to formally analyzing the computational effort and optimality of the proposed scheme, we thoroughly demonstrate its scalability and optimality in terms of reward in comparative experiments with a static rule-based approach as a baseline and a utility-driven approach using a constraint solver. These experiments are based on different failure profiles derived from real-world failure logs. We also investigate the impact of different failure profile characteristics on the scalability and reward to evaluate the robustness of the different approaches.
\end{abstract}



\authorsaddresses{Authors' addresses: S. Ghahremani and H. Giese, Hasso Plattner Institute, University of Potsdam, Prof.-Dr.-Helmert-Stra\ss e 2-3, Potsdam, D-14482, Germany; emails: \{sona.ghahremani, holger.giese\}@hpi.de; T. Vogel, Humboldt-Universit\"at zu Berlin, Unter den Linden 6, Berlin, D-10099, Germany; email: thomas.vogel@cs.hu-berlin.de.}

\maketitle
\renewcommand{\shortauthors}{S. Ghahremani et al.}

\section{Introduction}\label{sec:intro}

There are various ways of realizing self-adaptation adopting the MAPE-K feedback loop~\cite{Kephart&Chess2003} and in particular the analysis and planning phases.
On the one hand, \emph{rule-based} 

\newpage

\noindent
approaches~\cite{1537890,4061119} combine both phases. Adaptation is executed for specific events and under specific conditions by adaptation rules. In such approaches, events trigger the rules that subsequently check their conditions. If the conditions are fulfilled, the actions of the rules are applied and result in the envisioned changes. 
Thus, the applicable rules are identified (matched) and executed to adapt the system configuration at runtime.
The main strengths of such approaches are the readability, elegance, and the efficient processing of the rules. The drawbacks are 
the fact that the adaptation decisions are often only satisfying and the limited expressiveness of rules since rules typically just relate events to actions~\cite{1691383} without defining and performing any further computation for analysis and planning (e.g., to identify optimal actions).
On the other hand, \emph{utility-driven} approaches~\cite{Kephart+Walsh2004,Esfahani+2013} determine optimal adaptation decisions by using optimization techniques for planning that are guided by a utility function.
A utility function determines how valuable each possible system configuration is, and the optimization aims at identifying optimal configurations. However, the optimization usually prevents the approaches from scaling well for large configuration spaces. 
Scalability is further impeded by complex utility functions, as used in constraint solver-based approaches, so that mostly linear functions are used~\cite{1691383}.

Therefore, we present in this article a combined rule-based and utility-driven adaptation scheme that is scalable and guarantees optimal adaptation decisions with respect to the utility of the system and adaptation costs.
The combined approach achieves the individual benefits of both rule-based and utility-driven approaches, but it avoids the corresponding drawbacks with respect to the optimality of adaptation decisions and scalability.
Optimal adaptation decisions are achieved by selecting the best adaptation rules with respect to their impact on the overall utility and executing them in such a manner that rules with highest impact on the utility are prioritized. If adaptation rules have an equal impact on the overall utility, the ones with lower adaptation costs (i.e., the estimated time to execute a rule) are prioritized. This guarantees an optimal reward, which is gaining highest utility over time.
Scalability is achieved by an incremental approach that leverages events and patterns to efficiently identify adaptation issues and make adaptation decisions. Our approach is incremental, as its complexity is independent of the size of the system architecture and only influenced by the number of applicable adaptation rules and number of issues to be addressed by self-adaptation.

Our scheme particularly targets the architecture-based self-healing of large software systems---that is, resolving runtime failures by dynamically adapting the system architecture. For this purpose, we integrate our scheme in a \mbox{MAPE-K} feedback loop that operates on a causally connected runtime model of the system's runtime architecture.
Such self-healing systems are usually characterized by restrictions (e.g., adaptation is only needed if failures occur) that we exploit to guarantee optimal adaptation decisions.
Achieving optimality for self-healing requires finding the optimal adaptation rule to resolve a single failure, and the optimal ordering of executing such rules when multiple failures must be resolved at the same time. 
Whereas the former guarantees that each individual failure is handled by its best adaptation rule in terms of utility increase, the latter guarantees that rules achieving a larger increase of the utility or the same utility increase faster are executed first. Thus, the scheme achieves optimality in terms of the final utility achieved after adaptation and the utility over time (reward) gained during and after adaptation.
To achieve optimality, we use our former work to define the utility function in a pattern-based way for large dynamic architectures~\cite{Ghahremani+16}, and we further define the adaptation rules in a pattern-based way.
This joint use of patterns allows us to combine the utility and adaptation rules, and therefore to compute the impact of each rule application on the utility.
Based on these impact values for the rules and the knowledge about the estimated costs (execution time) of applying each rule, we can incrementally and efficiently determine and execute at runtime the optimal sequence of optimal adaptation~rules.

We demonstrate these benefits of our adaptation scheme by comparing it to two alternative solutions in simulations of \mRUBiS~\cite{2018-mRUBiS}. We show that our scheme is only slightly slower but reaches a higher utility over time (reward) than a static rule-based solution.
We further demonstrate that our scheme always makes optimal adaptation decisions similar to an alternative solution using a constraint solver. However, our scheme requires considerably less time than the solver, especially for large architectures.
As our approach is incremental, it faces less overhead and therefore scales better. As argued by~\citeN{Ghezzi2012}, incremental solutions are highly desirable for self-adaptive systems. In our earlier work~\cite{Vogel+2009,VogelNHGB10,VG10}, we presented an incremental scheme for the monitoring and execution phases of the feedback loop operating on architectural runtime models. The results of this work complement these earlier results by enabling the incremental analysis and planning with architectural runtime models, adaptation rules, and utility functions. Therefore, we focus in this article on the analysis and planning phases of the feedback loop.  

The idea of applying utility to pre-defined adaptation strategies or rules has been formerly practiced. For instance, RAINBOW~\cite{2012ChengStitch} employs utility theory to rank strategies taking into consideration their expected costs and benefits but without investigating scalable and timely adaptation decisions when facing large architectures and a multitude of issues (e.g., failures) to be addressed. We further distinguish our work from RAINBOW based on our incremental detection of multiple runtime failures and pattern-based runtime computation of the impact of the adaptation rules on the overall utility. Thus, we propose a scalable solution that guarantees to find the optimal adaptation decisions for large architectures in a timely manner, which maximizes the system utility over time~(reward).

The presented work extends our previous work~\cite{Ghahremani+17} with the following novel contributions.
First, we present formal algorithms for the analysis and planning phases of our self-healing scheme and a formal discussion of their computational effort and optimality.
Second, we strengthen the evaluation by using \textit{realistic} failure profile models that are based on \textit{real-world} data and that differ in scale and volatility to evaluate the scalability and optimality (reward) of our scheme in real-world settings. In this context, we also extend the evaluation from single to multiple MAPE-K runs. 
Third, we evaluate the robustness of our scheme by investigating the impact of different characteristics of failure profile models such as failure group size (FGS), inter-arrival time (IAT), and failure density on the scalability and optimality of the scheme.
Fourth, we discuss and justify assumptions of our scheme thoroughly with respect to optimality. Moreover, we analytically discuss and report on novel experiments of how potential violations of the assumptions impact the scalability and optimality of the scheme.
Finally, we improve the work by providing a more detailed discussion of the threats to validity and of related work with respect to timeliness and optimality of adaptation decisions.

The rest of the article is structured as follows.
We introduce architectural self-adaptation with runtime models and the pattern-based definition of utility in Section~\ref{sec:prerequisites}.
We detail our approach with its general scheme and its application in a feedback loop in Sections~\ref{sec:scheme} and~\ref{subsec:rulebased}.
We formally discuss the computational effort, optimality, and assumptions of our approach in Section~\ref{sec:analysis}.
In Section~\ref{sec:evaluation}, we evaluate our approach by comparing it to two self-healing approaches with respect to scalability and optimality, and we discuss threats to validity. The evaluation uses synthetic and realistic failure profile models for single and multiple MAPE-K runs, and also considers the violations of assumptions.
Finally, we review related work in Section~\ref{sec:related} and conclude with an outlook on future work in Section~\ref{sec:conclusion}.

\section{Prerequisites}\label{sec:prerequisites}

\subsection{Architectural Self-Adaptation and Runtime Models }
\label{subsec:runtime-model}\label{sec:arch-based}

To realize self-adaptation, a software system is equipped with a \textit{MAPE-K} feedback loop that \underline{m}onitors and \underline{a}nalyzes the system and, if needed, \underline{p}lans and \underline{e}xecutes an adaptation of the system, which is all based on \underline{k}nowledge~\cite{Kephart&Chess2003}.
In this context, many researchers consider the \textit{software architecture} as an appropriate abstraction level (e.g.,~\citeN{Oreizy+1999,Garlan+2009}) because self-adaptation can be \textit{generally} achieved by adding, removing, and reconfiguring components as well as connectors among components in the system~\cite{MageeKramer1996}.
For this purpose, the feedback loop maintains a \textit{runtime model} as part of its knowledge to represent the architecture of the system. This model is \textit{causally connected} to the system---that is, any relevant change of the system is reflected in the model and vice versa~\cite{Blair+2009}.
Thus, the MAPE phases operate on the runtime model to perform self-adaptation.
Moreover, a runtime model allows these phases to use model-driven engineering (MDE) techniques \cite{France+Rumpe2007}.
In earlier work~\cite{VG10,Vogel+2009,VogelNHGB10}, we presented incremental monitoring and execution phases that use MDE techniques and runtime models and that are the basis for this~work.

\begin{figure}[t]
	\centering
	\includegraphics[scale=.65]{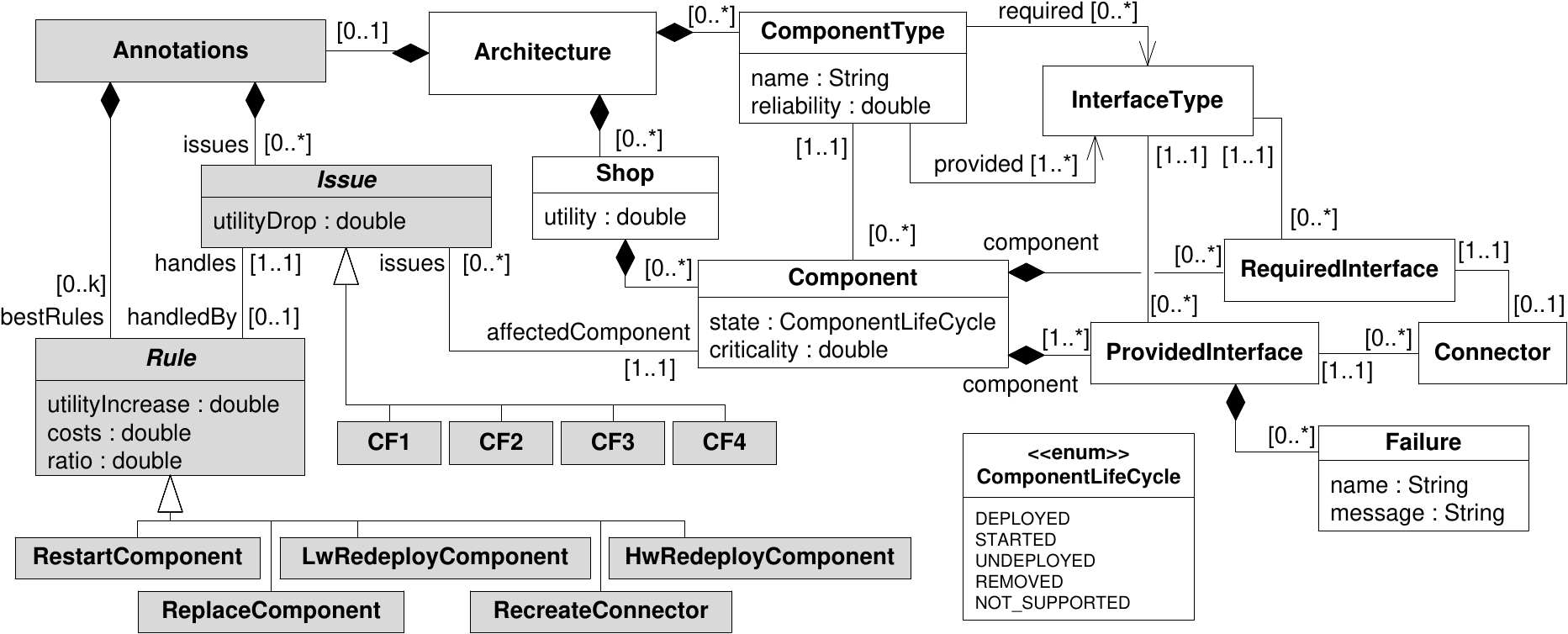}
	\caption{Simplified metamodel of the architectural runtime model.}
	\label{fig:metamodel}
	\vspace{-1em}
\end{figure}

As the running example, we use \textit{\mRUBiS}---an online marketplace that hosts an arbitrary number of shops, each consisting of 18 components~\cite{2018-mRUBiS}. Each shop can be configured differently and runs isolated from the other shops.
We are particularly interested in \textit{self-healing} to automatically repair runtime failures by architectural self-adaptation. This allows us to consider \textit{general} repair rules that adapt the architectural configuration of \mRUBiS.
Therefore, we equip \mRUBiS with a \mbox{MAPE-K} feedback loop that uses an architectural runtime model of \mRUBiS. Specifically, the model represents the runtime architecture of \mRUBiS according to the deployment of \mRUBiS in an  application server. For this purpose, the metamodel of the runtime model captures the \mRUBiS \elem{Architecture} with a set of \elem{ComponentType}s that require and provide \elem{InterfaceType}s (Figure~\ref{fig:metamodel}). For each \elem{Shop}, the same component types are instantiated to \elem{Component}s with their \elem{Provided-} and \elem{RequiredInterface}s. A \elem{Connector} links a required and a provided interface if both are of the same \elem{InterfaceType}. Using a \elem{ProvidedInterface} of a component may result in \elem{Failure}s in terms of exceptions. The \elem{ComponentLifeCycle} defines the \elem{state} of a \elem{Component}. 
These elements allow us to describe the runtime architecture of \mRUBiS and the occurred exceptions. The elements colored gray are relevant for self-adaptation and described later.

Using (meta)models and MDE techniques, we realize the analysis rules with model queries and the adaptation rules with in-place model transformations. For self-healing, the analysis rules query the architectural runtime model to identify issues such as failures in \mRUBiS. The adaptation rules determine how to modify the model and thus how to adapt the architecture to repair these issues.
To specify a model query, we use a pattern $P$ of a set of patterns $\mathcal{P}$ describing a structural fragment of the architecture $G$. Since the architecture is represented by the runtime model, we also use $G$ to refer to the model. 
An occurrence of a pattern $P$ in the model $G$ corresponds to a match $m$ of $P$ in $G$ (we write $G \models_m P$). For instance, a match identifies a failure in the architecture.
An adaptation rule $r$ in the rule set $\Re$ uses such patterns or already identified matches to localize a failure and to change the model in-place to repair the failure.
The automated matching of a pattern and the subsequent repair constitute the self-healing.

\subsection{Pattern-Based Architectural Utility}\label{sec:rbasedutility}\label{subsec:utility}\label{subsec:L-utility}

A \emph{utility} function $U$ is an objective policy that expresses how well each configuration of the system in its domain satisfies the functional and non-functional \emph{goals} of the system.  
For this purpose, $U$ assigns a real-value scalar desirability belonging to $[-\infty,+\infty]$ to any possible architectural system configuration $G$. 
Such scalar values allow us to compare different architectural configurations and to select the one with the highest utility as the best adaptation decision. 
Furthermore, the \textit{reward}, which is the accumulated utility over time, supports comparisons over time.

Defining a valid utility function is of high importance in an optimization problem of finding the best configuration since it is always the utility function and not the real utility of the system that is maximized. 
There has been extensive research on \emph {utility-driven} decision-making policies and elicitation of user preferences (e.g.,~\citeN{1317482}). 
A typical approach for architectural configurations is to compute for each non-functional property (e.g.,~reliability) the impact of alternative components providing similar functionality at different quality levels on the overall goals. A normalized linear utility function computes the weighted sum of these impact values over all properties given a concrete architecture with concrete alternatives selected. The weights represent the preferences of the user/developer, and the result is the utility of the given architecture~\cite{1128711}.
Such an approach can be used for planning self-adaptation to identify the target architecture, to which the system should be adapted.
Moreover, defining such utility functions is particularly challenging for large and dynamic architectures~\cite{Cheng2006}.

In the following, we outline our proposal to define utility functions for large, dynamic architectures based on patterns~\cite{Ghahremani+16}.
Due to the pattern-based utility definition, our utility functions can cope with dynamic architectural changes. 
We know that for a utility function evaluating an architectural runtime models must hold that (i)~the optimal architectural configuration where all of the system goals are optimally fulfilled must gain the maximum utility and that (ii)~if any goal becomes violated, this must lead to a decrease of the latest utility. 

According to (i), we include the impact of \emph{present architectural fragments} in the utility. 
We define such fragments by \emph{positive architectural utility patterns} $\Pa^+ = \{ P^+_1, \dots, P^+_k \}$ and capture their impact on the utility by utility sub-functions $U_i$.
Fragments defined by such patterns can target single or multiple components, and they can be both generic and component specific.
The impact of a pattern defined by $U_i$ may vary for each individual occurrence of the pattern in the architecture depending on the specific context of the present components. Thus, $U_i$ takes the context into account.

\begin{figure}[t]
	\begin{minipage}[t]{0.35\linewidth}
		\centerline{\includegraphics[scale=\scalefactor]{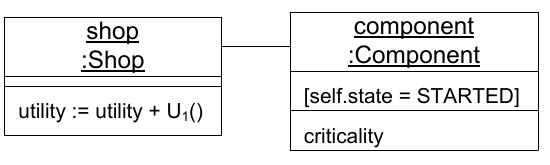}}
		\vspace{-0.5em}
		\caption{Positive architectural utility pattern $P^+_1$.}
		\label{fig:Pospatt}
	\end{minipage}%
	\hspace{.7em}%
	\begin{minipage}[t]{0.63\linewidth}
		\centerline{\includegraphics[scale=\scalefactor]{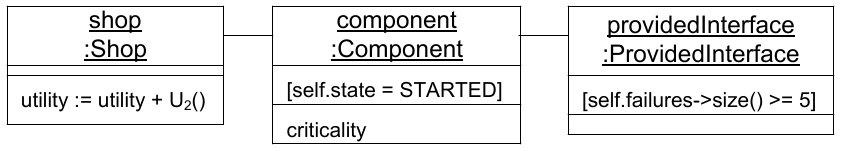}}
		\vspace{-0.5em}
		\caption{Negative architectural utility pattern $P^-_2$.}
		\label{fig:Antipatt}
	\end{minipage}%
	\vspace{-1em}
\end{figure}

As an example, Figure~\ref{fig:Pospatt} shows the positive pattern $P^+_1$ and the related utility sub-function $U_1$. This pattern conforms to the metamodel shown in Figure~\ref{fig:metamodel} and prescribes a started component that is associated with a shop and therefore contributes to the shop's functionality. Thus, this pattern targets a single component and is generic as it refers to any started component.
When matching this pattern for one component in the runtime model, the utility of the associated shop increases by $U_1$.
We define $U_1(G,m):=$ criticality of the component $\times$ reliability of the corresponding component type $\times$ connectivity of the component. If we match all components of a shop, the utility of the shop is the sum of the corresponding sub-utilities $U_1(G,m)$ for all of these components. Finally, the pattern is applied to all shops of \mRUBiS to obtain the utility for each shop.
Concerning the parts of $U_1(G,m)$, each component has a \elem{criticality} (cf.~Figure~\ref{fig:metamodel}) denoting its relevance for a shop. For instance, the \elem{Authentication} is more critical than the \elem{Reputation} component since the former is necessarily required by a shop to close a deal, whereas the latter is not.
Additionally, each component type has a \elem{reliability} (cf.~Figure~\ref{fig:metamodel}). For certain functionalities, alternative component types with different reliabilities exist (e.g., local vs. various third-party authentication services). Hence, selecting the most reliable alternative results in a higher utility increase.
The connectivity of a component as the number of associated \elem{Connector}s indicates the importance and accordingly influences the utility of the component. 
Thus, the pattern $P^+_1$ with its utility sub-function $U_1$ determines the context of a matched component in terms of criticality, reliability, and connectivity, which all influence the impact of an occurrence of the pattern on the utility.
In general, any data observable at runtime and represented in the runtime model can serve as the context of an architectural fragment (component) and be used for computing the impact on utility. At runtime, the context of a matched fragment is dynamically obtained from the runtime model when evaluating $U_i$.

According to (ii), we include the negative impact of undesirable situations defined by \emph{negative architectural utility patterns} $\Pa^- = \{ P^-_{k+1}, \dots, P^-_n \}$ in the utility. These patterns negatively affect the architecture such that they decrease the overall utility according to their utility sub-functions $U_i$. 
Examples of such negative patterns are occurrences of failures (e.g., exceptions).
As before, the impact may vary for each individual occurrence of a negative pattern depending on the specific context in the architecture. 
As an example, Figure~\ref{fig:Antipatt} shows the negative architectural utility pattern $P^-_2$ for \mRUBiS, which describes the case when five or more failures in terms of exceptions were thrown by a started component. 
Each occurrence of this negative pattern decreases the utility of the associated shop by $U_2$. We define $U_2(G,m) := -U_1(G,m)$ so that $U_2$ is negative.

\begin{figure}[t]
	\begin{centering}
		\includegraphics[scale=.75]{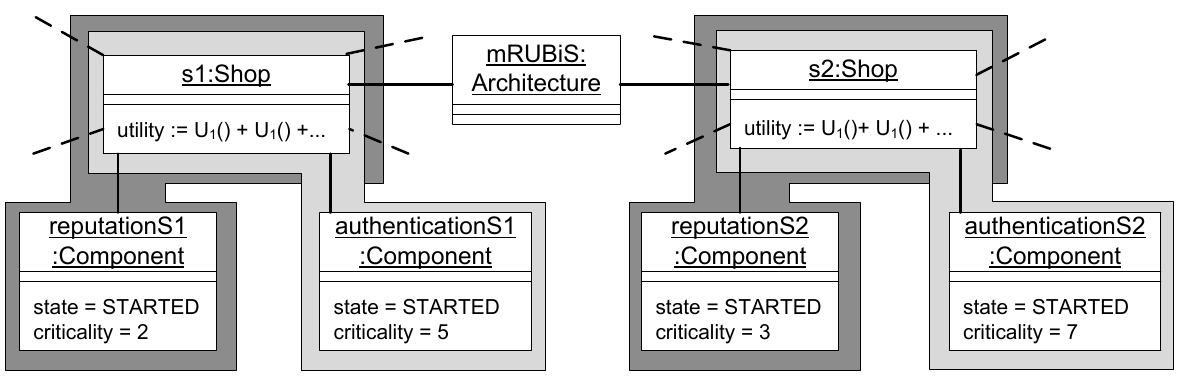}
		\vspace{-0.5em}
		\caption{Excerpt of the architectural runtime model with two matches of $P^+_1$ for each shop.}
		\label{fig:Utility}
	\end{centering}
	\vspace{-1em}
\end{figure}

Consequently, the positive patterns capture the possible utility gained by the current architectural configuration, whereas the negative patterns represent whether this potential is actually realized. If it is not realized, negative patterns occur in the architecture and correspondingly decrease the utility.
In \mRUBiS, adding a new shop with its components corresponds to a positive pattern that increases the utility. The number of the available shops and the relevant attributes of their comprising components~(e.g.,~criticality, connectivity) dynamically determine the overall utility of the architecture.
Occurrences of five or more failures (exceptions) in a component, component crashes, component removals, and connector crashes (cf.~Figure~\ref{fig:metamodel}) are examples of negative patterns that reduce the utility.
Figure~\ref{fig:Utility} shows an excerpt of the architectural runtime model with two matches of the positive pattern $P^+_1$ (cf.~Figure~\ref{fig:Pospatt}) for each of the two shops---that is, with two started components in each shop.
These two matches of the positive pattern $P^+_1$ are highlighted by different shades of gray for each shop. For instance, the elements \elem{s1} and \elem{reputationS1} denote one match, and \elem{s1} and \elem{authenticationS1} denote the other match of $P^+_1$ in shop \elem{s1}.
Each match increases the utility of the shop by $U_1(G,m)$ taking the characteristics of the specific component into account (e.g., the different criticality values of \elem{reputationS1} and \elem{authenticationS1}). The utility of a shop is the sum of $U_1(G,m)$ for all components of the shop, whereas the utility of the whole system is the sum of the utilities of all shops. 
Similarly, matches for negative patterns decrease the utility of the shops and thus of the system (not illustrated~in~Figure~\ref{fig:Utility}).

Therefore, considering $M_i(G) = \{ m\, |\, G\, \models_m P_i \}$ as the set of matches for the pattern $P_i$ in the current architectural configuration $G$, the overall utility function $U(G)$ accumulates all effects due to the matches of all $n$ patterns $\Pa = \{ P_1, \dots, P_n \}$\footnote{If we do not have to distinguish between positive and negative patterns, we omit the superscript $+$ and $-$ for the patterns $P \in \Pa$.}:

\begin{equation}\label{eq:utility}
U(G) 
:=
\sum_{i=1}^{n}
\sum_{m \in M_i(G)}
\!\!\!\!\!U_i(G,m)
\end{equation}

Analysis and adaptation rules can refer to such patterns. On the one hand, rules should identify and repair occurrences of negative patterns in the architecture. On the other hand, they should not affect existing positive patterns but rather enable new occurrences of positive patterns by repairing occurrences of negative patterns.

When matching a pattern, the concrete context is dynamically identified for each match in the runtime model. The context corresponds to a fraction of the runtime model that is navigated to obtain the required information to calculate $U_i$ at runtime. The definition of the pattern-based utility takes the context into account. Each pattern $P_i$ specifies a context that influences the utility sub-function $U_i$ and thus the increase or decrease of utility. 
For instance, each of the patterns $P^+_1$ (cf.~Figure~\ref{fig:Pospatt}) and $P^-_2$ (cf.~Figure~\ref{fig:Antipatt}) specifies the criticality of the component and the associated shop as the context. 
This context could be extended, for instance, by taking the component type into account so that the pattern would only match to components of a certain type (e.g., components realizing the authentication or user reputation).

Finally, the individual context of each occurrence of a pattern could cause variations in the utility after adaptation.
In our example of $P^+_1$ and $P^-_2$, consider an occurrence of $P^-_2$ with an absolute decrease in the utility of $|U_2|$. An adaptation resolving this occurrence of $P^-_2$ and enabling an occurrence of $P^+_1$ could achieve an increase of the utility $U_1$ that is larger than $|U_2|$. This is the case if the adaptation, for instance, replaces the faulty component with an alternative, more reliable component since the reliability of a component is a factor effecting the utility (see definitions of $U_1$ and $U_2$).

\section{Utility-Driven Rule-Based Adaptation Scheme}\label{sec:scheme}

We propose a utility-driven scheme to evaluate large, dynamic software architectures. As discussed in Section~\ref{subsec:utility}, utility functions map each architectural configuration of a software system to a scalar value indicating how well the configuration satisfies the goals. The need for evaluating \textit{dynamic} architectures is motivated by architectural self-adaptation. If adaptation is required, the feedback loop has to identify a suitable or even the optimal target configuration and select the adaptation rules that move the system to this configuration. For this purpose, a feedback loop can use the proposed scheme.
With this scheme, we are particularly interested in self-healing---that is, the automatic repair of runtime failures by \textit{general} rules that perform architectural adaptation and reconfiguration.

In this context, we express \emph{issues} (i.e., runtime failures) for an architecture as model \emph{patterns} such that concrete issues with different impacts on the overall utility $U(G)$ relate to occurrences (matches) of these patterns in the runtime model~$G$. 
Additionally, we can express an adaptation rule $r=(P,\dots)$ that is applied on the runtime model if the condition described by a model pattern $P$ is satisfied---that is, for each match of $P$ in the model. We denote for $r=(P,\dots)$ that a match $m$ for $P$ in the model $G_i$ exists and that applying the rule results in a modified model $G_j$ by $G_i \rightarrow_{r,m} G_j$.

\begin{figure}[t]
	\centering
	\includegraphics[width=.35\linewidth]{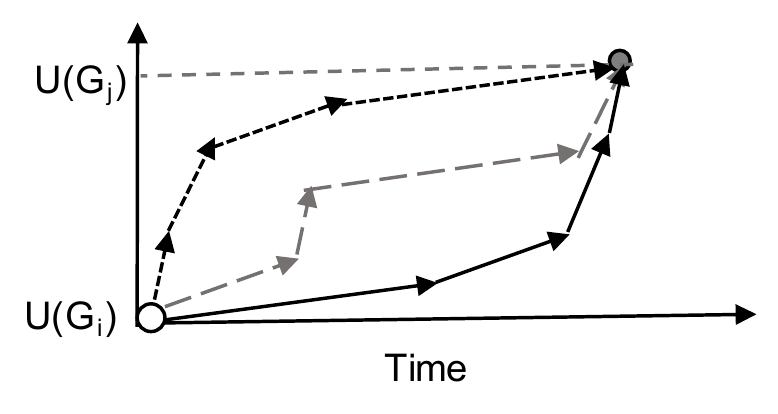}
	\vspace{-1em}
	\caption{Target configuration and different paths.}
	\label{fig:target}
\end{figure}

Our scheme can be mapped to a MAPE-K feedback loop operating on the runtime model. 
The \textit{monitoring} phase observes the current system configuration and updates the model. 
During \textit{analysis} and \textit{planning}, the scheme requires two decisions: the first decision is the target configuration of the system, and the second one is the rules and their matches that move the system to the target. 
These two decisions are inspired by the idea of model-predictive control that first defines a target and then predicts the optimal path to reach the target~\cite{Seborg+2011}. This is illustrated in Figure~\ref{fig:target} showing one target with three alternative paths to reach the target.

Considering self-healing, selecting an architecture where issues are repaired is equivalent to defining the target configuration. During the repair, selecting the best sequence of adaptation rules and their matches that resolve all issues is equivalent to building the path toward the target.
Paths that achieve earlier a larger increase of the utility are preferred. 
Finally, the last step of the feedback loop \textit{executes} these rules for their matches on the running system.

For a target configuration $G_j$, it must hold that its utility $U(G_j)$ must be higher than or equal to the utility $U(G_{i'})$ of all possible next and intermediate configurations $G_{i'}$ that are the outcomes of resolving issues in the current faulty configuration $G_i$. For self-healing, the target $G_j$ is always reachable unless there are resource limitations. 
To avoid enumerating the complete search space (i.e., all configurations reachable from $G_i$), our scheme computes the impact of each possible rule application for a match on the related utility sub-function and thus on the overall utility~($U(G_j)-U(G_i)$).
After defining the target~$G_j$, a set of adaptation rules with their matches has to be selected to reach~$G_j$. A sequence of rule applications changes the configuration $G_i$ toward $G_j$, which we denote as $G_i \rightarrow_{r_{i'},m_{i'}} G_{i'} \rightarrow ...\rightarrow_{r_j,m_j} G_j$.
Based on the impact of each rule application on the utility, we determine the path. To resolve a single issue, alternative rules are applicable and an estimation of their impacts on the utility allows us to select a conflict-free subset of them. We assume that for all such sets, we can compute the utility impacts regardless of the order in which the rules are executed. Our scheme is capable of doing so, since we assume that the impacts of the adaptation rules on the utility are independent of each another.
Our scheme guarantees that
(i) executing the selected rules eventually leads to the target $G_j$ with utility $U(G_j)$ and 
(ii) executing them in the right order results in the highest achievable reward (utility accumulated over time as represented by the area under curve in Figure~\ref{fig:target}).
To fulfill~(i), when there are two or more alternative rules to resolve the same issue, the scheme selects the rule with the highest impact on the utility. 
To achieve~(ii), the selected rules are executed in a decreasing order of their impact on the utility. 
We claim and will show that our scheme is optimal regarding the final utility $U(G_j)$ and the achieved reward.
To maximize the reward (utility over time), our scheme offsets the designated utility increase with the estimated time of executing each rule.
Reward optimality in the context of (i) is achieved by selecting the best rule in terms of utility increase for each issue. In the context of (ii), the approach is optimal since it prioritizes those rules from the selected rules that have larger impact on utility. In both cases, if multiple rules have the same impact on utility, faster rules are selected or prioritized as time~(besides utility) also affects the~reward.

\section{Realizing the Adaptation Scheme in a Feedback Loop}\label{subsec:rulebased}

The utility functions for runtime architectures as defined in Section~\ref{subsec:utility} allow us to follow an optimization-based approach that searches the configuration space and computes the utility for each possible configuration. However, such a solution for making adaptation decisions does not scale if the utility is computed for each configuration completely anew.
In contrast, the proposed utility-driven and rule-based scheme determines at runtime the \textit{impact} of each possible rule application on the utility.
Based on these impacts, it selects the optimal adaptation rule for each issue and identifies the optimal sequence of the rules for execution to maximize the reward.

\begin{figure}[t]
	\begin{centering}
		\includegraphics[width=.99\linewidth]{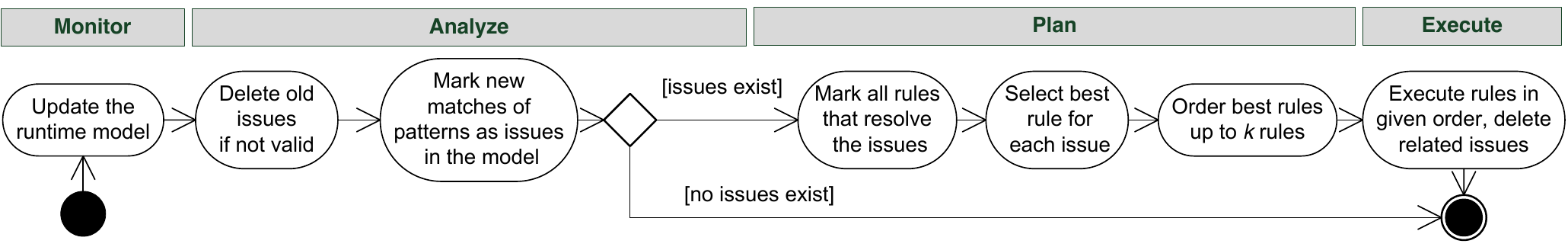}
		\caption{Different phases of the MAPE-K feedback loop realizing the proposed self-healing scheme.}
		\label{fig:MAPE}
	\end{centering}
\end{figure}

This scheme is realized by a \mbox{MAPE-K} feedback loop as shown in Figure~\ref{fig:MAPE}.
The execution of the feedback loop is triggered by every event that notifies about changes of the system under adaptation. The feedback loop is not reentrant so that all events that occur during a feedback loop run are queued. When the current feedback loop run is finished, the next run is triggered if there is at least one queued event, and this next run processes all of the currently queued events. Thus, there is no static frequency determined manually for executing the feedback loop.	
All of the four \mbox{MAPE} activities operate on the architectural runtime model.
During monitoring, the model is updated to reflect changes of the system.
The analysis deletes the old issues (i.e., matches of negative patterns that have been identified by a previous run of the loop) that are not valid anymore from the model. In addition, it detects the new issues and marks them in the model. 
The subsequent planning considers all possible adaptation rules that can address the existing issues. For each applicable rule, the impact on the utility and cost of execution are calculated. For each issue, the best rule regarding this impact and cost is selected. The selected rules over all issues are sorted according to their utility impact and cost, and stored in the model.
In the execution phase, the sorted list of rules is executed on the model and thus to the running system.
In the following, we provide a detailed description of the \mbox{MAPE-K} feedback loop activities realizing our scheme.

\subsection{Monitor}\label{subsec:m}

During monitoring, change events emitted by the system are processed and reflected in the runtime model.
Thus, the model is updated to represent the current architectural system configuration (cf.~\citeN{Vogel+2009,VogelNHGB10}). 
In our example, we observe the lifecycle \elem{state} of a component (e.g., to monitor whether a component has stopped, crashed, or been removed) and \elem{failure}s such as exceptions that occur when using a \elem{ProvidedInterface} (cf. metamodel in~Figure~\ref{fig:metamodel}).

\subsection{Analyze}\label{subsec:a}

In this phase, the updated runtime model is analyzed to check whether known matches of negative patterns (i.e., old issues) are still valid; otherwise, the annotations representing these issues are removed from the model. Moreover, the updated runtime model is analyzed to detect new matches of negative patterns (i.e., new issues) that enrich the known set of matches. This analysis is driven by events notifying about model updates (cf.~event-property-change mechanism).

As a first step, we compute the utility incrementally rather than for each configuration anew.
Given a former runtime model $G$ and an updated version $G'$, the set of new matches for the utility pattern $P_i$ is $M_i^{new} = M_i(G') \backslash M_i(G)$. Similarly, $M_i^{del} = M_i(G) \setminus M_i(G')$ contains the matches for the pattern $P_i$ that are no longer valid. 
We can therefore calculate the corresponding change of the overall utility $U(G') - U(G)$ by the utility change function $U_\Delta(G',G)$: 
\begin{equation}\label{equation:incr}
-
\sum_{i=1}^{n}
\sum_{m \in\!M^{del}_i}
\!\!\!\!\!U_i(G,m)
+
\sum_{i=1}^{n}
\sum_{m \in\!M^{new}_i}
\!\!\!\!\!U_i(G',m)
\end{equation}

\begin{figure}[t]
	\centering
	\includegraphics[scale=\scalefactor]{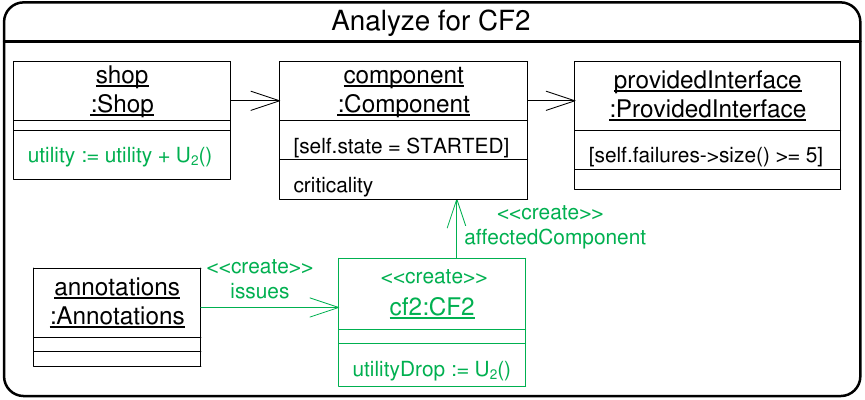}
	\caption{Annotating an occurrence (match) of a negative pattern in the runtime model.}
	\label{fig:A}
\end{figure}

Besides computing the change of the utility, we keep track of the identified matches for negative patterns (the issues). 
In the case of self-healing, all of the architectural utility patterns that need to be matched and resolved are the negative patterns $\Pa^- = \{ P^-_{k+1}, \dots, P^-_n \}$.
For this purpose, the analysis phase adds \elem{Annotations} to the runtime model. It checks the model for occurrences of negative patterns, which are then annotated in the model as \elem{Issue}s pointing to the \elem{affectedComponent}.
We consider the following issues:
crashes (\elem{CF1}) and removals (\elem{CF3}) of components, 
occurrences of \elem{Failure}s in terms of exceptions (\elem{CF2}), 
and connector crashes (\elem{CF4}) (cf.~Figure~\ref{fig:metamodel}).
For instance, Figure~\ref{fig:A} shows an analysis rule realized by a \textit{story pattern}~\cite{FNTZ98_ag} that detects the negative pattern $P^-_2$ (cf. Figure~\ref{fig:Antipatt}). The occurrence of this pattern results in a drop in the shop's utility by $U_2$. This rule \elem{create}s the \elem{CF2} annotation with the computed \elem{utilityDrop} that points to the affected component.
Here, we omit the details to avoid multiple annotations for the same~issue. 

\subsection{Plan}\label{subsec:p}

Based on the annotations representing new or remaining matches for negative patterns and thus issues in the runtime model, our approach incrementally proceeds during the planning by
(1)~computing the set of all possible adaptation rule applications (matches) that resolve each issue,
(2)~selecting the best rule application for each issue based on the impact on utility and cost, and
(3)~finally ordering the best rule applications across all issues to minimize the lost reward.

\subsubsection{Compute All Possible Adaptation Rule Matches.}

In self-healing, applying an adaptation rule always leads to an improved utility as it resolves an issue that has previously caused a drop in the utility.
For this case, we will show that adaptation rules have to be linked to negative patterns. Moreover, knowing the matches for these patterns allows us to incrementally compute all relevant adaptation rule matches---that is, all rules that are applicable to resolve the issues.

For any self-healing adaptation scheme with pattern-based utility functions, all of the patterns that need to be matched and resolved are the negative ones and the following observations must hold:
(1)~if there are no matches of negative patterns, there is no need for adaptation and no improvement of the utility is possible, and
(2)~any improvement of the utility must necessarily resolve identified matches of negative patterns, as otherwise no improvement is possible. 

Thus, we can safely assume that (A1)~for any adaptation rule $r_j = (P_j, \dots)$ in the set of all adaptation rules $\Re$, it must hold that a negative pattern $P^-_i$ exists such that any match $m_j$ for $P_j$ ($m_j$ makes $r_j$ applicable) includes a match $m_i$ for $P^-_i$ ($m_i$ denotes an occurrence of the negative pattern $P^-_i$). Thus, any adaptation rule must be linked to a negative pattern such that the rule can only be applied if there is an occurrence of the negative pattern. Otherwise, the rule could be applied even though there is no occurrence of a negative pattern, and no utility improvement can be achieved, which contradicts observation (1).
It can be the case that the pattern $P_j$ of $r_j$ has a larger context and is thus more restricted than the negative pattern $P^-_i$. However, both patterns are exactly the same in the presented examples.

Furthermore, we can plausibly assume that (A2)~for rule $r_j = (P_j, \dots)$ in $\Re$ and any match~$m_j$ for $P_j$ with the included match $m_i$ for the related negative pattern $P^-_i$, it holds that applying $r_j$ for $m_j$ will make the match $m_i$ invalid. 
This means that executing an applicable rule resolves the related occurrence of the negative pattern by repairing the issue.
Otherwise, $r_j$ does not handle the identified occurrence of the negative pattern $P^-_i$ and therefore does not lead to the improvement of the utility as expected by observation (2).
To keep our considerations simple, we consider the case where each rule covers exactly one negative pattern. 
Based on these assumptions, we can compute all matches for rules incrementally given the set of new matches $M_i^{new}$ for the related negative pattern $P^-_i$ by the analysis phase.

\begin{figure}[t]
	\begin{centering}
		\includegraphics[scale=\scalefactor]{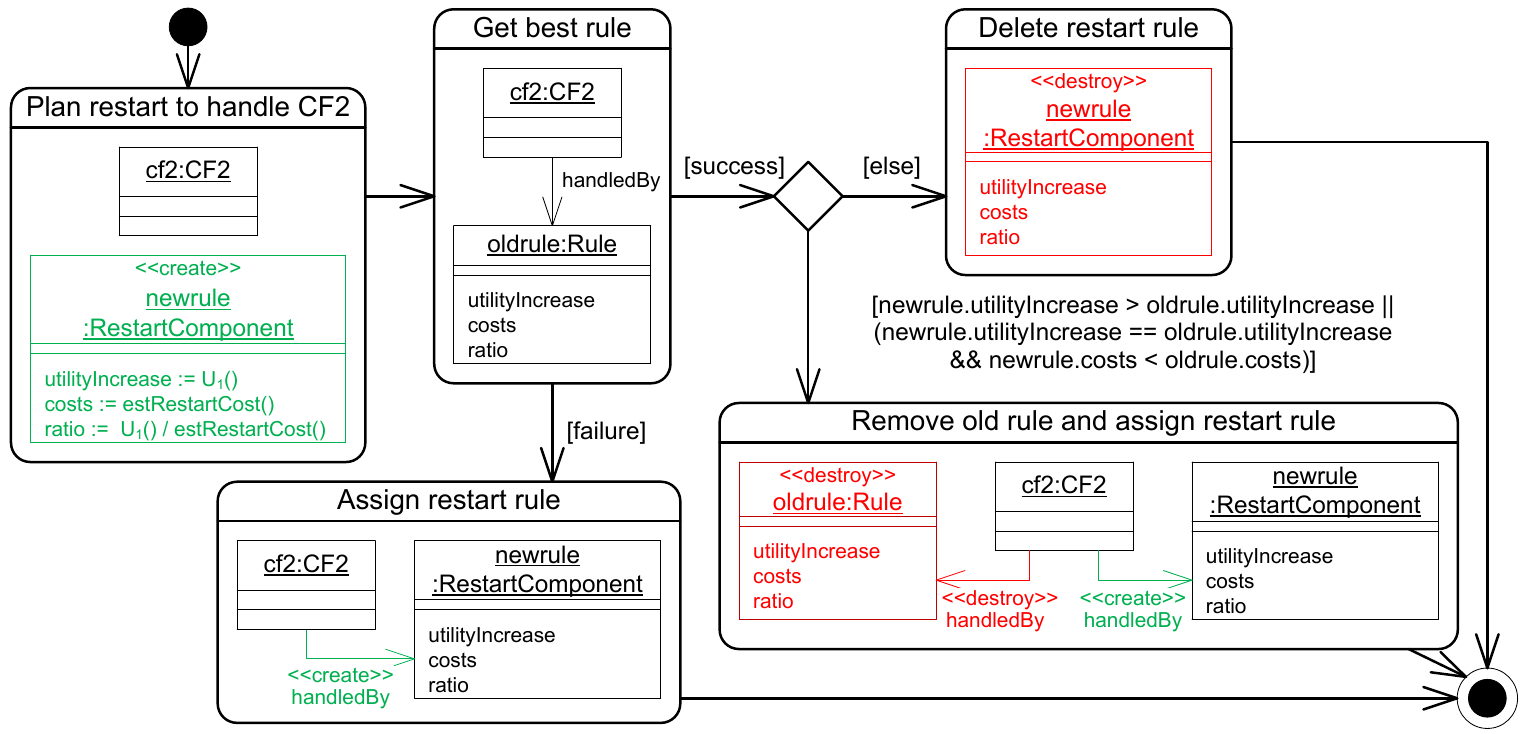}
		\caption{Rule for planning an adaptation.}
		\label{fig:P}
	\end{centering}
\end{figure}

In general, performing an adaptation with MAPE-K consists of a planning and an execution part. The planning decides which adaptation rules among all possible ones should be applied. The execution part actually applies the selected rules to prescribe an adaptation in the runtime model that is subsequently propagated to the system (cf. causal connection in Section~\ref{subsec:runtime-model}).
In our case, the planning selects the adaptation rules to be executed by enriching the model with \elem{Rule} annotations that will \elem{handle} the identified \elem{Issue}s (cf.~Figure~\ref{fig:metamodel}). These rules are finally enacted by the execution phase. 
As adaptation rules, we support restarting, redeploying, and replacing components, as well as recreating connectors (cf.~Figure~\ref{fig:metamodel}). For the redeployment, there are two variants. The light-weight variant keeps the latest configuration, whereas the heavy-weight variant adapts the configuration (e.g., parameters) of the redeployed component.
To realize the planning rules, we use story diagrams as shown in Figure~\ref{fig:P}. Using story diagrams, we can structure story patterns (nodes of the rule) in a control flow that complies with UML activity diagrams (cf.~\citeN{FNTZ98_ag}).
Considering Figure~\ref{fig:P}, the first node of the planning rule matches the \elem{CF2} annotation created by the analysis phase, and it creates the new adaptation rule of type \elem{RestartComponent} as a \elem{Rule} annotation to repair the \elem{CF2} instance. Such a planning rule exists for each adaptation rule (e.g., component restart) and for each issue (e.g., \elem{CF2}) to which the adaptation rule is applicable.
The decision for the adaptation rules to be executed is made by selecting the best among all applicable rules for each~issue.
 
\subsubsection{Select the Best Adaptation Rule Match for Each Negative Pattern Match.} 

Each issue in terms of a negative pattern match can be repaired by alternative adaptation rules. Thus, the planning must select one rule for execution. To determine the \textit{best} among all matched adaptation rules for each issue, we compute the impact on the utility and the costs of executing each of these rules. 
Formally, for a single rule $r_j = (P_j,\dots)$ where $P_j$ extends the negative pattern $P^-_i$, it holds that each time $r_j$ is applied to $m_j$, the match $m_i$ for a negative pattern $P^-_i$ is removed (see A2). We further assume that~(A3) $r_j$ does not result in any new match or removed matches besides $m_i$ for any negative pattern. 
Then, we conclude for any $G,\,G'$ resulting from applying rule $r_j$ to $G$ for match $m_j$ ($G \to_{r_j,m_j} G')$ that:

\begin{equation}\label{eq:delta}
U^{r_j}_\Delta(G,m_j)
:=
U_\Delta(G',G) 
=
U_i(G,m_i) 
\end{equation}

This way, we can compute \textit{locally} the impact of adaptation rules on the utility if the assumptions (A1) through (A3) hold.
If further assumption (A4) holds that rule $r_j$ does not affect any utility sub-function for any match $m_k$ of another negative pattern $P^-_k$, then applying $r_j$ for a match $m_j$ does not affect the impact on the utility of any other rule $r_k$ for match $m_k$. Thus, if (A1) through~(A4) hold, we can \textit{independently and locally} compute the utility impact of each rule. 
However, there can be cases where the side effect of applying a rule~$r_j$ (i.e., $G \to_{r_j,m_j} G'$) results in new matches for one or more positive patterns. In such cases, the impact on the utility by the corresponding positive utility sub-function of these matches is added to $U^{r_j}_\Delta(G,m_j)$ in Equation~(\ref{eq:delta}). 
For this reason, it must hold that all of the potential positive patterns are completely within the scope of the application condition and side effect of $r_j$ and do not match only partially~(A5). 
Otherwise, matches for the positive patterns cannot be enabled by applying $r_j$.
Thus, the impact can be considered in $U^{r_j}_\Delta(G,m_j)$ because the resulting formula for the corresponding increase of the utility can be determined at development time. 
An example for such a case is when replacing the local authentication component of \mRUBiS with a third-party service while each available service results in a different increase on the utility depending on the \elem{reliability} of the service.
Similarly, the costs in terms of execution time for adaptation rules can be estimated by a cost function $Cost^{r_j}(G,m_j)$ for each application of a rule. Thus, the estimated execution time may depend on the match $m_j$ with its context in $G$. 

Hence, for each issue, the planning rules determine the expected \elem{utilityIncrease} $U^{r_j}_\Delta(G,m_j)$ and \elem{costs} $Cost^{r_j}(G,m_j)$ of executing each applicable adaptation rule to the system, and they select the best among all applicable adaptation rules.
For instance, Figure~\ref{fig:P} shows a planning rule for repairing a \elem{CF2} issue by restarting the affected component. 
Following the analysis phase capturing matches of negative patterns with annotations such as \elem{CF2} (cf.~Figure~\ref{fig:A}), all planning rules of $\Re$ are checked if they match such annotations. The ones that match are able to plan how an issue should be handled by enriching the model with annotations for adaptation rules. 
In the example in Figure~\ref{fig:P}, the first node (story pattern) of the planning rule matches the \elem{CF2} annotation and creates the \elem{Rule} annotation of type \elem{RestartComponent}. 
It further determines the \elem{utilityIncrease} and \elem{costs} of restarting the component (see attributes of \elem{newrule:RestartComponent}). 
For this purpose, the utility sub-function $U_1$---as discussed in Section~\ref{subsec:utility}---is used to calculate the expected \elem{utilityIncrease} and thus the positive impact of executing the rule on the utility.
This utility sub-function takes the context of the match and thus runtime information into account (e.g., the reliability and criticality of the affected component).
Moreover, cost functions such as \elem{estRestartCost()} for each adaptation rule type estimate the \elem{costs} of executing the rule such as restarting a component in the system. In our work, the cost estimation for each rule type is static, context independent, and based on past measurements of the time that is needed to execute the corresponding change to a running system. However, the cost functions can be more elaborate, taking the context of the match into account (e.g., it is more costly to replace a component with a higher connectivity, i.e., a larger number of associated connectors).
By using the \elem{utilityIncrease} and \elem{costs} of all applicable adaptation rules for an issue, our scheme selects the best rule.
For this reason, the planning checks for each applicable adaptation rule (e.g.,~\elem{newrule:RestartComponent} in Figure~\ref{fig:P}) whether this new rule results in a higher \elem{utilityIncrease} than the rule that has been determined as the best one so far within this run of the feedback loop (see \elem{oldrule:Rule} in Figure~\ref{fig:P}). In the case that both rules achieve an equal \elem{utilityIncrease}, the approach checks if the new rule has lower \elem{costs}.
If this old rule (i.e.,~the rule that has been determined as the best rule so far) does not exist, the planning selects the new rule to handle the \elem{CF2} issue (see story pattern \elem{Assign restart rule}). Otherwise, if the new rule is better than an already selected rule (i.e.,~the old rule) with respect to \elem{utilityIncrease} and \elem{costs}, the old rule~is~deleted and the new rule is selected (see story pattern \elem{Remove old rule and assign restart rule} in Figure~\ref{fig:P}). For the case when the old rule is better than the new rule, the planning proceeds with the old rule and deletes the annotation for the new rule (see story pattern \elem{Delete restart rule}). 
Thus, the planning rules select for each issue the best adaptation rule that is going to be executed and that is associated to the issue by the \elem{handles}/\elem{handledBy} association in the runtime model (cf.~Figure~\ref{fig:metamodel}).

\subsubsection{Order the Execution of All Selected Adaptation Rule Matches.}

The final planning step determines the order in which the issues should be resolved if multiple issues occur at the same time. We assume that $k$ issues could be repaired and thus $k$ adaptation rules could be executed within one cycle of the feedback loop.
Therefore, the $k$ best adaptation rules (see the previous planning step) are sorted in descending order regarding their impact on the overall utility divided by the costs. This metric combining the benefits and costs of an adaptation rule is reflected by the \elem{ratio} attribute of a \elem{Rule} (see Figure~\ref{fig:metamodel}) and computed by the planning rules (see Figure~\ref{fig:P}).
Applying the adaptation rules in this order, as maintained by the association \elem{Annotations.bestRules} in the runtime model (see Figure~\ref{fig:metamodel}), guarantees in the execution phase that the maximal utility is re-established as fast as possible and that the loss of reward is minimized.
As depicted in Figure~\ref{fig:target}, prioritizing the adaptation rules with a higher impact on utility (i.e.,~rules with higher slopes) maximizes the area under the curve (a curve being a path to the target configuration) that is recognized as utility over time or reward.

\subsection{Execute}\label{subsec:e}

\begin{figure}[t]
	\begin{centering}
		\includegraphics[scale=\scalefactor]{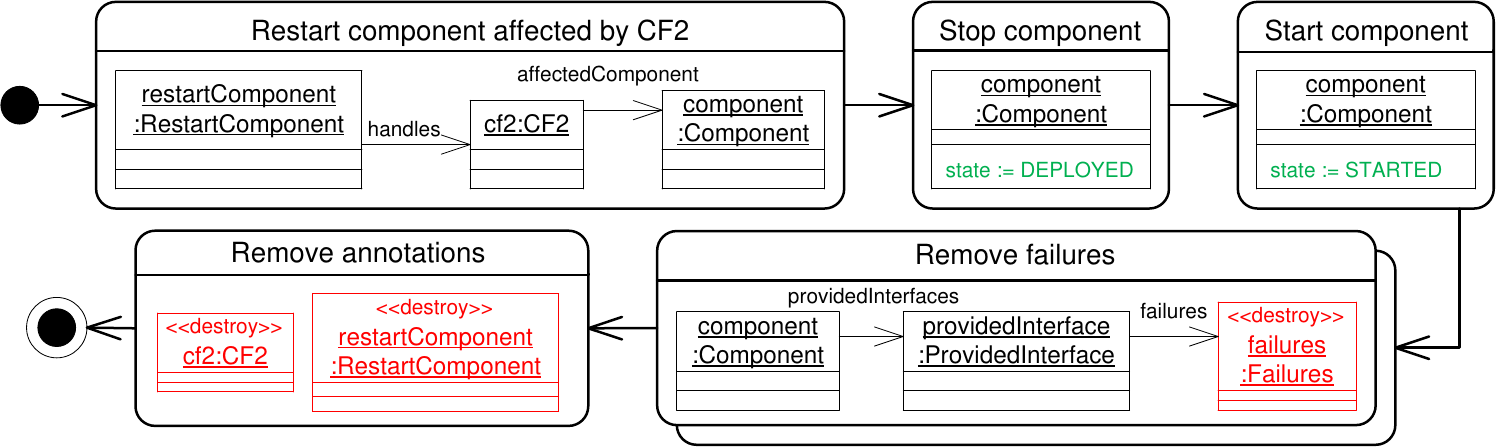}
		\caption{Rule for executing a component restart.}
		\label{fig:E}
	\end{centering}
\end{figure}

Given the sorted list of adaptation rules from the planning phase, this phase executes these rules accordingly in a sequential manner.
Thus, each issue is handled with the most appropriate rule, and the rules are executed in the order such that those with the best trade-off (\elem{ratio}) of \elem{utilityIncrease} and \elem{costs} are executed first.
Similar to monitoring, this phase follows an incremental scheme in executing adaptation rules on the runtime model and propagating the corresponding changes through the causal connection to the running system~\cite{VG10}. 
Figure~\ref{fig:E} illustrates an adaptation rule for executing a restart of a component to address \elem{CF2}.
Based on the analysis and planning phases (see Figures~\ref{fig:A} and~\ref{fig:P}), an adaptation rule, in this case \elem{restartComponent}, has been selected to handle \elem{CF2} affecting the specific \elem{component} (see the first node in Figure~\ref{fig:E}). This component is then restarted by setting its lifecycle state to \elem{DEPLOYED} (i.e., stopped) and then to \elem{STARTED} (see the second and third nodes). After that, the runtime model is cleaned by removing (\elem{destroy}ing) the observed exceptions (\elem{failures}) and the annotations for the executed rule (\elem{restartComponent}) and the handled  issue (\elem{CF2}).

\section{Analysis and Discussion}\label{sec:analysis}

We now analyze and discuss the computational effort and the optimality of the utility and reward for our scheme. We will present the generic algorithms for the analysis and planning and show that their computation is done incrementally and the resulting adaptation leads to an optimal reward. Finally, we discuss limitations concerning our assumptions.

\subsection{Detailed Algorithms for the Analysis and Planning}\label{subsec:algorithms}

The $analyze()$ and $plan()$ algorithms of our approach use a global data structure (a runtime model) defined by the metamodel in Figure~\ref{fig:metamodel}. 
The singleton object $annotations$ of type $Annotations$ captures the current matches for issues with its $issues$ association of type $Issue$.  $Annotations$ also captures the best $k$ matches for rules to repair the issues with its $bestRules$ association of type $Rule$. Both can be accessed by $annotations.getAllIssues()$ and $annotations.getAllBestRules()$. 
Additionally, the association $handledBy$ between the matches for issues and matches for rules maintains the best rule match for each issue. 
For an $Issue$, we assume that a test $check()$ exists that checks in constant time whether the match is still valid.
To maintain matches for issues in a global list without double entries, we assume constant-time operations $annotations.existsIssue()$, $annotations.addIssue()$, and $annotations.deleteIssue()$. This is possible when using index data structures with unique matches as index.
Similarly, we require constant-time operations $annotations.addBestRule()$ and $annotations.resetBestRule()$ to maintain and clear a list of the best $k$ matches for rules with respect to their $ratio$. 
Here, $k$ is chosen large enough to cover as many rule applications as possible to fit into the time window of a single MAPE-K run. 
The dedicated time window of a MAPE-K run is a design decision. A shorter window allows more frequent planning. As a result, rules achieving a high impact on the utility are executed earlier and not blocked by rules that achieve lower increases but are already scheduled to be executed by a previous MAPE-K run. This is beneficial if the planning is time efficient. For a selected time window, $k$ can be determined using estimates of planning time of the scheme and rule execution time. Thus, $k$ can vary for different time windows. Since at most $k$ adaptation rules are applied in a single MAPE-K run, $k$ is a constant upper bound on the rule elements that have to be stored and ordered in the runtime model.
A run of the feedback loop is finished after the $k^{th}$ adaptation rule has been executed. The remaining failures that have not been addressed in this run will be handled in the subsequent run of the feedback loop.

\begin{algorithm}[t]
	\caption{$analyze(C)$ }\label{algo:analyze}
	    \ForAll(\tcp*[f]{ iterate over all old issues of $c$ }\nllabel{algo-analyze-begin-all-changes-issues}){ $issue \in annotations.getAllIssues()$ }{%
            \If(\tcp*[f]{ delete old issue if no longer valid }){ ! $issue.check()$ }{ 
                     $annotations.deleteIssue(issue)$ \tcp*{ delete issue from global list }     
            }	    
      }\nllabel{algo-analyze-end-all-changes-issues}      
  \ForAll(\tcp*[f]{ iterate over all modified or created elements in the model}\nllabel{algo-analyze-begin-all-changes}){ $c \in C$ }{%
      \ForAll(\tcp*[f]{ consider all relevant patterns $P_i$}\nllabel{algo-analyze-begin-all-pattern}){ $P_i \in \Pa \mbox{ containing a node that may be matched to } c$ }{
          \ForAll(\tcp*[f]{ find all new issues for $P_i$ }\nllabel{algo-analyze-begin-all-pattern-issues}){ $m: G \models_{m} P_i \land m \mbox{ contains } c$ }{
             \If(\tcp*[f]{ check if issue for $P_i$ and $m$ already exists }\nllabel{algo-analyze-begin-all-pattern-issues-check}){ $! annotations.existsIssue(P_i, m)$ }{ 
                  $annotations.addIssues($new $Issue(P_i, m))$            \tcp*{ add new issue to the global list }\nllabel{algo-analyze-all-pattern-issues-add}
              }
          }\nllabel{algo-analyze-end-all-pattern-issues}
      }\nllabel{algo-analyze-end-all-pattern}
 }\nllabel{algo-analyze-end-all-changes}
\end{algorithm}

\begin{sloppypar} 
Given a set $C$ of elements in the runtime model that have been changed by the monitoring, the $analyze()$ procedure shown in Algorithm~\ref{algo:analyze}
(1) removes any old issue stored in $annotations.getAllIssues()$ that are no longer valid (lines~\ref{algo-analyze-begin-all-changes-issues}--\ref{algo-analyze-end-all-changes-issues}) and
then
(2) iterates over all changes $ c \in C$ (lines~\ref{algo-analyze-begin-all-changes}--\ref{algo-analyze-end-all-changes}) for all patterns in $\Pa$ relevant for $c$ (lines~\ref{algo-analyze-begin-all-pattern}--\ref{algo-analyze-end-all-pattern}) and for all possible new issues whose matches contain $c$ (lines~\ref{algo-analyze-begin-all-pattern-issues}--\ref{algo-analyze-end-all-pattern-issues}) to check whether these issues are new (line~\ref{algo-analyze-begin-all-pattern-issues-check}) so that they are added to the global list (line~\ref{algo-analyze-all-pattern-issues-add}). 
Thus, after executing $analyze(C)$, the set of issues stored in $annotations.issues$ is up-to-date.
\end{sloppypar}

\begin{algorithm}[b]
	\caption{$plan()$  }\label{algo:plan}
	{ annotations.resetBestRules()  \tcp*{ erase all entries form the list of best rules and matches }\nllabel{algo-plan-erase-best-ones}}
  \ForAll(\nllabel{algo-plan-begin-all-issues}){ issue $\in$ annotations.getAllIssues() }{
	    \ForAll(\nllabel{algo-plan-begin-all-issues-rules}){ $r_i \in \Re \mbox{ that can resolve } issue$ }{
  	        \ForAll(\nllabel{algo-plan-begin-all-issues-rules-matches}){ $m: G \models_{m} r_i$ with $m \supseteq issue.match()$ }{
              { rule := new Rule(m)} \tcp*{ create new rule with utilityIncrease and costs }
              { oldrule := issue.getHandeledBy()} \tcp*{ get rule with best  utilityIncrease and costs identified up to now }
              \If{ oldrule $=$ NULL || rule.utilityIncrease $>$ oldrule.utilityIncrease || (rule.utilityIncrease $==$ oldrule.utilityIncrease \&\& rule.utilityIncrease/rule.cost $>$ oldrule.utilityIncrease/oldrule.cost) }
                  { issue.setHandeledBy(rule)  \tcp*{ rule is better concerning utilityIncrease and costs than the old one }\nllabel{algo-plan-set-handledby}
              }
          }\nllabel{algo-plan-end-all-issues-rules-matches}
      }\nllabel{algo-plan-end-all-issues-rules}
      { rule := issue.getHandeledBy()} \tcp*{ get rule with the overall best utilityIncrease and costs }\nllabel{algo-analyze-get-best-one}
      { annotations.addBestRules(rule, rule.utilityIncrease/rule.cost)  \tcp*{ keep $k$ best rule for all issues according to the ratio }\nllabel{algo-plan-add-best-ones}}
  }\nllabel{algo-plan-end-all-issues}
\end{algorithm}

\begin{sloppypar} 
In addition, the $plan()$ procedure shown in Algorithm~\ref{algo:plan} checks 
for all current issues in $annotations.getAllssues()$ (lines~\ref{algo-plan-begin-all-issues}--\ref{algo-plan-end-all-issues}),
for all rules that may resolve the match of the issue (lines~\ref{algo-plan-begin-all-issues-rules}--\ref{algo-plan-end-all-issues-rules}), 
and for all matches of these rules that extend the match of the issue (lines~\ref{algo-plan-begin-all-issues-rules-matches}--\ref{algo-plan-end-all-issues-rules-matches}) 
whether the identified rule match is better than the currently best rule match captured by $issue.getHandledBy()$ (i.e., the utility increase is larger, or if the utility increase is the same, the costs are smaller). If so, the currently best rule match is replaced by the newly found rule match.
Thus, for each issue, the best rule match is determined and added to the global list of $k$ best rules in line~\ref{algo-plan-add-best-ones}. 
Note that this global list is initially erased in line~\ref{algo-plan-erase-best-ones}.
\end{sloppypar}

In our example, we exploit the fact that each issue is always linked to one unique component (cf.~Figure~\ref{fig:metamodel}) to realize the $annotations.existsIssue()$ test. The test just checks whether the same issue already exists for the related~component.

\subsection{Computational Effort for the Analysis and Planning}\label{subsec:computational-effort}

For the $analyze()$ and $plan()$ procedures in Algorithms~\ref{algo:analyze} and~\ref{algo:plan}, the patterns of the issues and rules do not need to be found by a global search, but by a local search starting from a change or an existing match. 
It is assumed that (1) patterns have a constant upper bound concerning their size and (2) the links for associations that have to be traversed by a local search for matches have a small constant upper bounds (assumption (A6)). As a result, finding a single match for an issue or rule requires only a constant computational effort. 
Based on this final additional assumption (A6) that typically holds, we will establish that the $analyze()$ and $plan()$ procedures only require an incremental computational effort in $O(\Delta + \Delta')$, where $\Delta'$ is the number of unprocessed issues and $\Delta$ is the number of changes in the  runtime model.

In the $analyze()$ procedure (Algorithm~\ref{algo:analyze}), the first main loop from line~\ref{algo-analyze-begin-all-changes-issues} to \ref{algo-analyze-end-all-changes-issues} requires $O(\Delta')$ steps to check for $\Delta'$  many old, unprocessed issues stored in $annotations.issues$ whether they are still valid or not. Conducting such a single check requires constant time.
The second main loop from line~\ref{algo-analyze-begin-all-changes} to \ref{algo-analyze-end-all-changes} iterates over all changes $c \in C$ with $|C|=\Delta$ and thus results in $O(\Delta)$ iterations.
The inner loop from line~\ref{algo-analyze-begin-all-pattern} to \ref{algo-analyze-end-all-pattern}, which considers all patterns in $\Pa$ potentially matching $c$, has a constant number of iterations because of the small numbers of patterns, a constant subset of which can actually be matched to $c$.
The other inner loop from line~\ref{algo-analyze-begin-all-pattern-issues} to \ref{algo-analyze-end-all-pattern-issues}, which considers all matches for such patterns that actually include $c$, is bounded by the number of changes $|C|$. Therefore, it requires $O(\Delta)$ iterations.
Thus, the $analyze()$ procedure requires an incremental computational effort in $O(\Delta+ \Delta')$, and the resulting list of current issues is in $O(\Delta + \Delta')$.

The $plan()$ procedure (Algorithm~\ref{algo:plan}) at first erases the list $annotations.bestRules()$ of $k$ best rule matches from the last cycle in constant time (see $annotations.resetBestRules()$ in line~\ref{algo-plan-erase-best-ones}). 
Then, up to $O(\Delta + \Delta')$ many current issues are handled in the loop from line~\ref{algo-plan-begin-all-issues} to \ref{algo-plan-end-all-issues}. 
For each iteration processing an issue, constant many rules that may resolve this issue are considered in the loop from line~\ref{algo-plan-begin-all-issues-rules} to \ref{algo-plan-end-all-issues-rules}.
In the worst case, the number of these rules is equal to $|\Re|$  ($O(1)$)---that is, the number of the adaptation rules in the finite set $\Re$. Thus, the number of the considered rules is limited by this constant upper bound.
As a result, constant many rule matches (each considered rule either matches or not) that extend the match of the issue are considered in the loop from line~\ref{algo-plan-begin-all-issues-rules-matches} to \ref{algo-plan-end-all-issues-rules-matches}.
While iterating through all applicable rules for an issue (bounded by $|\Re|$), one \emph{best} rule regarding the utility increase and cost is selected.
Therefore, only constant many times the best rule match for an issue is updated and stored in $issue.getHandledBy()$ in line~\ref{algo-plan-set-handledby}.

Once for each match of an issue, at the end of the loop from line~\ref{algo-plan-begin-all-issues} to \ref{algo-plan-end-all-issues}, \textit{the} best rule match to resolve this issue as stored in $issue.getHandledBy()$ (line~\ref{algo-analyze-get-best-one}) is considered in line~\ref{algo-plan-add-best-ones}.
Restricting the~number of issues that can be resolved within a feedback loop run to $k$, only $k$ and thus constant many best rule matches are kept with $annotations.addBestRule()$ (line~\ref{algo-plan-add-best-ones}). Thus, this step has a constant computational effort.
Consequently, the $plan()$ procedure requires an incremental computational effort in $O(\Delta+\Delta')$, and the resulting list of best rule matches contains $k$ elements ($O(1)$).

The monitoring and execution can be done event based and incrementally (cf.~\citeN{Vogel+2009,VogelNHGB10,VG10}). The $analyze()$ and $plan()$ procedures require incremental computational efforts. Overall, we conclude that the whole MAPE-K loop can operate in a highly efficient, incremental manner. The proposed scheme requires $O(\Delta + \Delta')$ steps with $\Delta'$ being the number of unprocessed issues and $\Delta$ the number of changes of the runtime model.

\subsection{Optimality of a Single MAPE-K Run}\label{subsec:optimality-of-single-run}

In this section, we discuss why the presented scheme guarantees an optimal adaptation behavior concerning utility and reward in a single MAPE-K run, given the assumptions we made and an appropriate selection of $k$~(cf.~Section~\ref{subsec:algorithms} for appropriate selection of $k$).

Executing all $k$ selected matches for adaptation rules in $annotations.bestRules$ guarantees a maximal increase of the overall utility because it removes all matches of negative patterns (assumption (A2)) and it does not affect any other matches for such patterns (assumptions (A3) and (A4)). Thus, the finally achieved utility is maximal after executing all selected rules. 
This effect on the overall utility remains in the system as long as the system is operating, and any other selection of rules that would lead to a lower utility will also result in a lower reward even though its costs might be lower. Thus, a faster adaptation achieving a lower final utility does not pay off since the system continues operating with a lower utility. Such a system will be affected by any future failures equally to the system after the optimal adaptation that operates at a higher utility level.
Furthermore, the ordering of the $k$ adaptation rules in $annotations.bestRules$ ensures for the time window when the rules are executed that the resulting reward is maximal. The reason is that any reordering of two rule matches with different $ratio$s results in a lower reward (i.e., smaller area under the curve in Figure~\ref{fig:target}). Any more complex reordering can be achieved by iteratively exchanging two rule matches, which would eventually also lead to a lower reward. Thus, the correctly ordered sequence results in the maximal reward.

Thus, when $k$ is larger or equal to the number of identified issues and all of these issues can be resolved within the time window of a single MAPE-K run, executing the $k$ rules in the given order results in the maximal reward. In this case, all identified issues can be repaired by one run of the feedback loop.
However, if $k$ is smaller than the number of identified issues, the resulting reward of executing the chosen sequence of adaptation rule matches is still optimal given the rationale for defining $k$~(cf. Section~\ref{subsec:algorithms}). This would avoid blocking the execution of rules achieving a high impact on the utility by rules that achieve lower increases but are already scheduled to be executed by a previous MAPE-K run.

\subsection{Discussion of the Assumptions}\label{subsec:discussions-for-assumptions}

In the following, we consider the assumptions we made (Table~\ref{tab1}) and discuss their justifications and the consequences if they do not hold. 
A violation of assumption (A1) indicates that the adaptation rules can be applied even if there is no match for a negative pattern. This can be ruled out for self-healing systems where we consider only rules that repair occurrences of negative patterns. However, it might be an issue for self-optimizing systems where adaptation rules are continuously applied to maximize the utility while a notion of negative patterns might not exist. 

\begin{table}[t]
	\begin{center}
		\caption{List of Assumptions}
		\label{tab1}
		\label{tab:assumptions}
		\begin{tabular}{ l p{12.4cm} }
			\toprule
			(A1) & For any in-place model transformation rule $r_j = (P_j, \dots)$ in the adaptation rule set $\Re$, it must hold that a negative pattern $P^-_i$ exists such that any match $m_j$ for $P_j$ includes a match $m_i$ for $P^-_i$. \\
			(A2) & For any rule $r_j = (P_j, \dots)$ in the rule set $\Re$ and any match $m_j$ for $P_j$ and the included match $m_i$ for the related negative pattern $P^-_i$, it holds that applying $r_j$ for $m_j$ will make the match $m_i$ for $P^-_i$ invalid. \\
			(A3) & Applying $r_j$ does not result in any new match or removed matches besides $m_i$ for any negative pattern. \\
			(A3a) & Applying $r_j$ does not result in any new match for any negative pattern. \\
			(A3b) & Applying $r_j$ does not result in any removed matches besides $m_j$ for any negative pattern. \\
			(A4) &  Applying $r_j$ does not affect any utility sub-function for any match $m_k$ for another negative pattern $P^-_k$, then applying a rule $r_j$ for a match $m_j$ does not affect the impact on the utility for any other rule $r_k$ and match $m_k$. \\
			(A5) &  All of the potential positive patterns are completely within the scope of the application condition and side effect of $r_j$ and do not match only partially. \\
			(A6) &  The links for associations that have to be traversed for local search of matches have always small constant upper bounds. 		\\ \bottomrule
		\end{tabular}
	\end{center}
\vspace{-1em}
\end{table}
 
Assumption (A2) states that any adaptation rule, if applied, is effective and therefore resolves the corresponding issue. A violation of (A2) implies that adaptation rules are not always effective---that is, applying a rule does not always resolve the issue. We can rule out this case given a deterministic behavior of adaptation rules.

Adaptation rules that cause new issues are not reasonable so that we can safely accept assumption~(A3a). We suggest designing the rules in such a way that they immediately resolve all additional issues they might cause. 
In the proposed scheme, if a rule accidentally causes a new match of a negative pattern, which will trigger another rule (violation of (A3a)), the new match will be detected and resolved in the next feedback loop run.
In contrast, a violation of assumption (A3b) results in a case where applying rules impacts the applicability of other rules. 
An example of such a violation is applying a rule that replaces a faulty component, which makes the repair rule of the related faulty connectors inapplicable, for instance, because the new version of the component needs different types of connectors and thus a different rule to re-establish the connectors.
In this case, the issue of the faulty component overlaps with the issue of the faulty connectors, and the issue of the faulty connectors will not be resolved in this but in the subsequent feedback loop run if it can be matched by a negative pattern.
However, we suggest designing the rules in a way that avoids such unwanted dependencies between rules. One way is to define the scopes of the rules such that each issue type is completely treated by one rule and the scopes do not overlap. This requires that the scopes of different types of issues do not overlap; otherwise, overlapping issue types should be combined to one type.
For the example, such a design results in a scope for the rule replacing a faulty component that also covers the related faulty connectors. 
Meanwhile, the scope for the connector repair rule would be restricted to faulty connectors that are not associated to faulty components. 

Assumption (A4) excludes cases where executing a rule affects the impact of executing another rule on the overall utility. A violation of (A4) implies dependencies between the rules similar to a violation of (A3).
Again we suggest designing the rules in a way that avoids such unwanted dependencies (cf. previous paragraph). 
If the adaptation rules influence each other regarding the impact on the utility (violation of~(A4)), our proposed scheme would not necessarily find the optimal rule for each issue and consequently the optimal ordering of all rules since it does not take such dependencies into account. However, all issues would still be resolved, although not necessarily with the ``best'' rules.

Assumption (A5) indicates that executing a repair rule achieves the intended improvement of the utility. If it does not hold, then the utility function or the context to calculate the impact of the rules on the utility is not appropriate. This requires a more expressive utility function or a larger context. 
Although it can be challenging to define an appropriate utility function (cf.~\cite{SGCAHG18}), establishing a larger context is in principle always possible by splitting such rules into multiple, more specialized rules that have a larger context to achieve the overlap with the positive~patterns. 

Assumption (A6) comprises the constant upper bound of the size of the patterns and the local search for matches of patterns/rules. This is justified based on the simple nature of the patterns that we encountered.
Nevertheless, if the assumption does not hold, other schemes such as pure rule-based ones might also have high execution costs.

To summarize, the assumptions listed in Table~\ref{tab1} are usually justified for rule-based self-healing approaches because rules that are not triggered by any issue or that do not resolve any issue do not make any sense (see (A1) and (A2)).
Rules that cause new issues are not reasonable and could thus be excluded (see (A3a)).
Rules that affect other issues or other rules are not useful and should thus be avoided (see (A3b) and (A4)). 
Rules that do not completely cover the positive patterns should be avoided (see (A5)).
Rules and patterns that are large or that do not allow a local search are not usual (see (A6)).
In this context, we can often assume a deterministic behavior of adaptation rules. However, there might be cases in which rules will not always succeed in repairing issues.  
We will therefore investigate such cases by considering a likelihood for the success of each adaptation rule in Section~\ref{subsec:violation-of-assumption-A2}.

\section{Evaluation}\label{sec:evaluation}

To evaluate our scheme, we use the \mRUBiS simulator~\cite{2018-mRUBiS}, a variant of RUBiS that is frequently used for validating self-adaptation targeting performance~\cite{Patikirikorala+2012}. 
\mRUBiS is a marketplace on which users sell or auction items. The simulator emulates the marketplace and provides fault injection capabilities. It emulates failures in \mRUBiS by reflecting them in the architectural runtime model as it would be otherwise done by monitoring the faulty system.
To determine the injected traces of failures, we use synthetic and realistic failure profile models.\footnote{All experiments and simulations were conducted on a machine with OS\,X $10.10$ and an Intel Core i5 2.6-GHz processor with 9 GB of memory. All measurements were conducted following benchmarking guidelines by~\citeN{Sestoft+2013}.}

Specifically, \mRUBiS can host different numbers of shops ($1$\,to\,$1\text{,}000$), each containing $18$ components with different \elem{criticality} and \elem{connectivity} values. The utility of a shop is the sum of the sub-utilities of all of the components in the shop. As described in Section~\ref{sec:arch-based}, we equipped \mRUBiS with a MAPE-K feedback loop. The three issues \elem{CF1}, \elem{CF2}, and \elem{CF3} are the negative patterns that affect the system. The rule set $\Re$ includes the adaptation rules, each representing a repair plan. Each rule has three attributes: \elem{costs}, \elem{utilityIncrease}, and \elem{ratio} (see Figure~\ref{fig:metamodel}). \elem{Costs} refers to the expected execution time of the rule, \elem{utilityIncrease} is the impact on the utility when applying the rule, and \elem{ratio} is the fraction of \elem{utilityIncrease}/\elem{costs}. 

In Section~\ref{subsec:single}, we validate the scalability and optimality of our scheme with analytical and real experiments using a \textit{synthetic} failure profile model to generate failure traces. In these experiments, we compare our scheme to two alternative self-healing approaches for a single MAPE-K run.
The evaluation is extended in Section~\ref{subsec:exteval} by looking at multiple MAPE-K runs and by using various \textit{realistic} failure profile models.
Experiments regarding the violation of the assumptions listed in Table~\ref{tab:assumptions} are discussed in Section~\ref{subsec:violation-of-assumption-A2}.
Finally, we discuss the threats to validity in Section~\ref{subsec:threats-to-validity}.

Overall, the evaluation compares three self-healing approaches: static, solver, and u-driven.

\paragraph{Static.}
This approach is purely rule based and uses static priorities for rules without any utility function. Thus, the \elem{costs} and \elem{utilityIncrease} of the repair rules are defined at design time so that a rule is selected statically for each~\elem{CF}. The \elem{utilityDrop} caused by each \elem{CF} is also estimated at design time leading to a fixed order in which the issues are resolved. 

\paragraph{Solver.}
This approach is purely utility based and uses the IBM ILOG CPLEX constraint solver~\cite{Cplex} for planning. Specifically, it uses the utility function described in Equation\,(\ref{eq:utility}) for the sequence of rule applications as its objective function. The tasks of assigning proper adaptation rules to each \elem{CF} and ordering them are defined as optimization problems. This approach maximizes the objective function as the overall utility of the system after each decision. 

\paragraph{U-driven.}
As described in Section~\ref{subsec:rulebased}, our approach computes the impact of different adaptation rules at runtime using the utility function shown in Equation\,(\ref{eq:utility}). It selects the rule with the largest impact on the overall utility and the lowest costs in the case of identical impacts. The order in which \elem{CF}s are addressed is determined by the \elem{ratio} of \elem{utilityIncrease}/\elem{costs}, which are all dynamically computed based on the runtime observations regarding the \elem{affectedComponent} of the \elem{CF}s.

\vspace{4pt}

Thus, the three approaches have different planning phases while they share the same incremental behavior---as suggested for our approach---for the other MAPE-K phases. For instance, the solver approach uses the constraint solver only to solve the planning problem of selecting the best repair rules and their order for execution. Possible ways of how to monitor and modify the system/runtime model are completely pre-defined and identical for all approaches. 

\subsection{Experiments for Single MAPE-K Runs}\label{subsec:single}

In the following, we conduct a set of analytical experiments that separately evaluate the two main steps of our approach for a single MAPE-K run: (1) selecting the best adaptation rules and (2) ordering them for execution. 
Then, we discuss experiments for different sizes of the \mRUBiS architecture to investigate the scalability and experiments with different failure traces extracted from a \textit{synthetic} failure profile model to investigate the optimality in terms of reward.

\subsubsection{Analytical Experiments.}\label{subsubsection:Analytic-single}

We consider \mRUBiS with 100 shops (1,800 components). The experiments start with occurrences of three failures of type \elem{CF1}, \elem{CF2}, and \elem{CF3} causing the utility of the system to drop. These utility drops are followed by a single MAPE-K run that resolves the three \elem{CF}s by executing three repair rules.
Here, we show that the u-driven approach makes the optimal decision during rule matching for \elem{CF}s by selecting the rule that results in the maximum increase in the overall utility. In contrast, the static approach fails to do so and hence is non-optimal. We also show how the order in which the adaptation rules are executed impacts the achieved reward. 

When a match for an issue is detected, our approach computes the \elem{utilityIncrease} and the \elem{costs} of all possible matches among the adaptation rules. The effect on the increase in the utility achieved by applying each rule remains in the system as long as the same component is not affected again by another issue. However, the costs of applying a rule has only a short-term effect on the overall utility. Cost defines when the expected increase of the utility can be realized. Thus, in our approach, rules with the highest \elem{utilityIncrease} are prior to those with lower increase but less \elem{costs}. The type of the occurred issue and the specific component that is affected by the issue determine the \elem{utilityIncrease} and \elem{costs}.

\begin{figure}[t]
	\centering
	\includegraphics[width=0.99\linewidth]{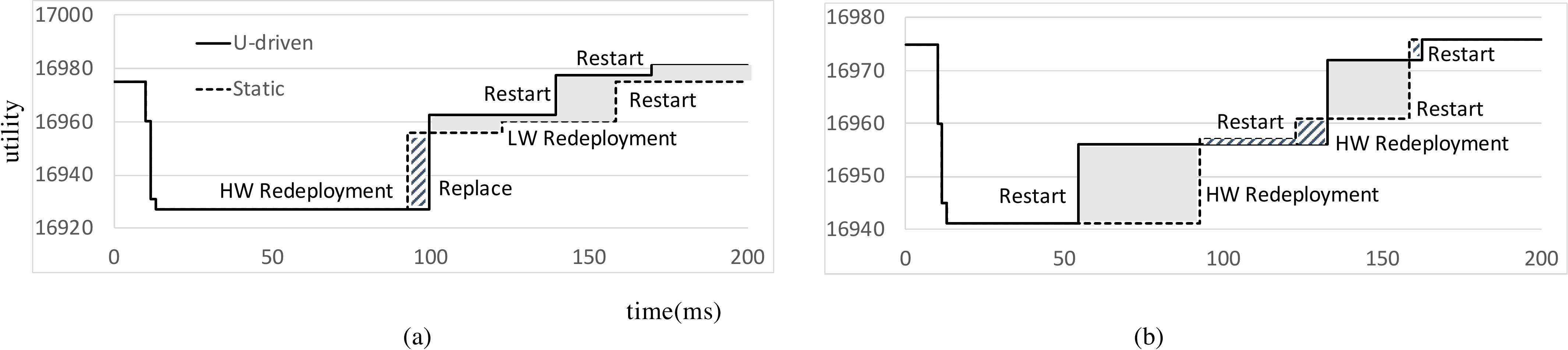}
	\vspace{-1em}
	\caption{Lost reward of the static approach compared to the u-driven approach due to non-optimal rule selection (a) and rule ordering (b).}
	\label{fig:analyticstatic}
\end{figure}

Figure~\ref{fig:analyticstatic}(a) shows a case where the static approach fails to reach the maximum utility due to non-optimal rule selection. 
Both the static and u-driven approaches select \elem{CF3} to be resolved first. The static approach performs a \elem{Heavy Weight (HW) Redeployment} while the u-driven approach \elem{Replace}s the affected component and reaches a higher utility. 
Here, the static approach selects a rule with less \elem{cost} and thus achieves the utility increase earlier than the u-driven approach (cf.~the hachure region), but it obtains a considerably smaller reward. The impact of this non-optimal rule selection remains in the system during the whole experiment and results in a lower reward for the static approach equal to the area of the gray-colored regions.
As the second decision, the static approach resolves \elem{CF1} by a \elem{Light Weight (LW) Redeployment} while the u-driven approach decides to resolve \elem{CF2} by a \elem{Restart} of the component that has a higher impact on the overall utility. 
As the last decision, the static approach resolves \elem{CF2} by a \elem{Restart} and reaches the same increase in utility as the u-driven approach in the second rule execution, but with a delay and thus with a lower reward as the utility is lower during the execution time. 
The u-driven scheme finishes the adaptation by repairing \elem{CF1} with a \elem{Restart} rule.
The static approach is slightly faster than the u-driven approach due to avoiding all of the runtime computations. The gray and hachured regions respectively represent the lost and gained reward of the static approach compared to the u-driven approach. The reward gained by the static approach due to less overhead and choosing the cheaper \elem{HW Redeployment} over the \elem{Replace} rule does not compensate for the loss of reward due to making non-optimal decisions. 

To back our claim for optimality of the u-driven approach, this approach executes the adaptation rules in the optimal order so that the maximum utility over time is achieved. We investigate this aspect in Figure~\ref{fig:analyticstatic}(b). The order in which our approach resolves the issues is such that those repair rules resulting in larger utility increase to cost ratio are prioritized. The static approach decides for the order at design time. This can be done considering the type of the issues. A reasonable order of rules based on the three issues in our example is
(1)~removals of components (\elem{CF3});
(2)~crashes of components that, however, might still be operating to a certain extent (\elem{CF1}); and
(3)~occurrences of \elem{Failure}s in terms of exceptions (\elem{CF2}).
This ordering fails to take into account the actual utility provided by the \elem{affectedComponent} that is a function of criticality, connectivity, and reliability (cf.~Section~\ref{subsec:utility}). 
These properties can dynamically change such that they are only known at runtime. Figure~\ref{fig:analyticstatic}(b) illustrates a case where the static approach fails to address the issues in the right order. 
Despite the fact that the static approach makes the optimal decision regarding the rule selection and that both the u-driven and static approaches achieve the same final utility, which is not necessarily always the case (cf.~Figure~\ref{fig:analyticstatic}(a)), the static approach loses reward due to the suboptimal ordering (gray regions) and gains only a slight improvement due to the lower overhead in planning time (hachured~region).

Considering Figure~\ref{fig:analyticstatic}(b), the u-driven approach first repairs \elem{CF2} and the static one \elem{CF3}. 
The static approach applies a \elem{HW Redeployment} reaching a slightly higher utility but with considerably larger costs than the u-driven approach that \elem{Restart}s the affected component.
Because of the large cost of the rule selected by the static approach, the static approach loses utility that is equal to the area of the first gray region. In contrast, it gains utility that is equal to the hachured area over the u-driven approach due to repairing a different issue first.
As the second repair decision, the static approach resolves \elem{CF1} by a \elem{Restart} while the u-driven approach resolves \elem{CF3} by a \elem{HW Redeployment}. 
Finally, the static approach resolves \elem{CF2} by a \elem{Restart} and reaches the same increase in utility as the u-driven approach but loses utility over time due to the suboptimal execution order. The u-driven approach saves the repair of \elem{CF1} by a \elem{Restart} for the last repair decision because in this scenario, repairing a \elem{CF1} has less impact than \elem{CF2} and \elem{CF3} on the utility. 

\begin{figure}[t]
	\begin{centering}
		\includegraphics[width=0.68\linewidth]{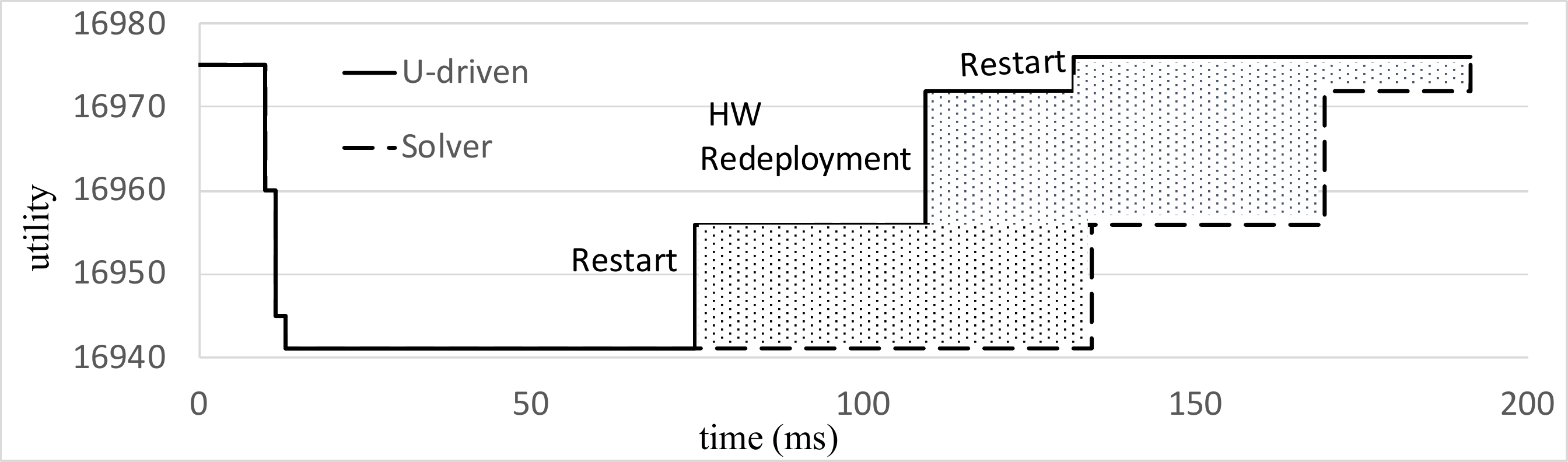}
		\caption{Lost reward of the solver approach compared to the u-driven approach due to longer planning time.}
		\label{fig:analyticsol}
	\end{centering}
	\vspace{-1em}
\end{figure}

We conducted the same experiments as in Figure~\ref{fig:analyticstatic}(b) to compare the solver and u-driven approaches.
Both approaches make identical decisions regarding the rule matching and ordering of the adaptation rules so that both reach the optimal configuration and achieve the same final utility (Figure~\ref{fig:analyticsol}). However, the solver approach achieves it after a considerable delay due to its computational planning overhead, which depends on the size of the architecture and number of the issues.  
Despite the fact that both approaches select  each time the same rules with identical \elem{costs} and \elem{utilityIncrease}, and decide on the same ordering of the rules, the planning overhead of the solver approach causes a delay that results in the lost reward compared to the the u-driven approach (see the dotted area in Figure~\ref{fig:analyticsol}).

\subsubsection{Experiment Design.}\label{subsubsec:simpleSC}

The following experiments for scalability and reward share the same experiment design. We use a \emph{synthetic} failure profile model to generate \emph{synthetic} failure traces. These traces are synthetic because the distribution of the failure profile model is not based on real-world data. 
Particularly, we generate four different failure traces. Each trace has a fixed \textit{failure group size} (FGS)---that is, the number of failures occurring before each MAPE-K run (cf.~\citeN{Gallet2010}), of either $1$, $10$, $100$, or $1\text{,}000$. 
Moreover, the different \elem{CF} types of the failures are equally distributed for each trace.
The \textit{failure exposure time} (FET) is the time window within which the failures belonging to the same FGS (i.e., before one MAPE-K run) occur. For an FGS larger than $1$, we assume that all of the failures of this group occur at once---that is, FET is only an instant and therefore is considered as $0$.
The \textit{inter arrival time} (IAT) is the time between occurrences of two groups of failures. In the synthetic failure profile model, we assume a long enough IAT between the MAPE-K runs such that all of the current failures are resolved before the failures of the next group occur. 
The \textit{failure density} of each trace is the overall number of failures injected by the trace. For example, considering four MAPE-K runs, the failure density of a trace with FGS of $10$ (i.e., $10$ failures occur before each MAPE-K run) is $40$.
  
\subsubsection{Experiments for Scalability.}\label{subsubsec:Eresult}

To compare the planning time and scalability of the three self-healing approaches, we tested them on \mRUBiS with four different sizes of the architecture ($18$, $180$, $1\text{,}800$, and $18\text{,}000$ components) and with four synthetic failure traces (FGS of $1$, $10$, $100$, and $1\text{,}000$ failures).
For each experiment, we consider single MAPE-K runs, one self-healing approach, one size of the architecture, and one trace to measure the planning time of the approach.
Each experiment is repeated $300$ times or until the standard deviation of the planning time is below $5\%$. The measurements follow benchmarking guidelines~\cite{Sestoft+2013}, and we report the mean of the results. 
For the experiments, we only consider meaningful combinations of architecture size and number of failures. Thus, we do not inject a large number failures to small architectures. Therefore, we do not present any data where more than $10$ ($100$) failures occur in a system with $18$ ($180$) components.
Since we are interested in the planning phases of the three self-healing approaches that have the same monitoring, analysis, and execution phases, we only present the data for the planning phases in~Table~\ref{fig:Table}.

\begin{table}[t]\centering
	\ra{1.2}
	\caption {Planning Time of the Three Self-Healing Approaches (in milliseconds)}
	\vspace{-1em}
	\label{fig:Table}
	\scalebox{.77}{
		\begin{tabular}{r|rrr|rrr|rrr|rrr}\toprule
			& \multicolumn{3}{c|}{$1$ failure} & \multicolumn{3}{c|}{$10$ failures} & \multicolumn{3}{c|}{$100$ failures} & \multicolumn{3}{c}{$1\text{,}000$ failures} \\
			\# Comp. & Static & U-driven & Solver & Static & U-driven & Solver & Static & U-driven & Solver & Static & U-driven & Solver \\ \cmidrule{1-13}
			18 & 0.76 & 0.89 & 5.02 & 10.37 & 14.36 & 55.68 & -- & -- & -- & -- & -- & -- \\
			180 & 0.68 & 0.89 & 5.01 & 9.71 & 13.58 & 59.07 & 14.22 & 17.70 & 219.54 & -- & -- & -- \\
			1,800 & 0.61 & 0.74 & 4.83 & 10.60 & 13.47 & 58.24 & 13.82 & 26.65 & 211.09 & 54.50 & 60.09 & 3,216.60 \\
			18,000 & 0.65 & 0.71 & 4.90 & 10.14 & 13.87 & 71.93 & 21.80 & 26.38 & 271.51 & 127.80 & 171.31 & 3,611.95 \\
			\bottomrule
		\end{tabular}
	}
\end{table}

\begin{figure}[t]
\centering
\begin{minipage}[t]{.5\textwidth}
  \centering
  \includegraphics[width=0.9\linewidth]{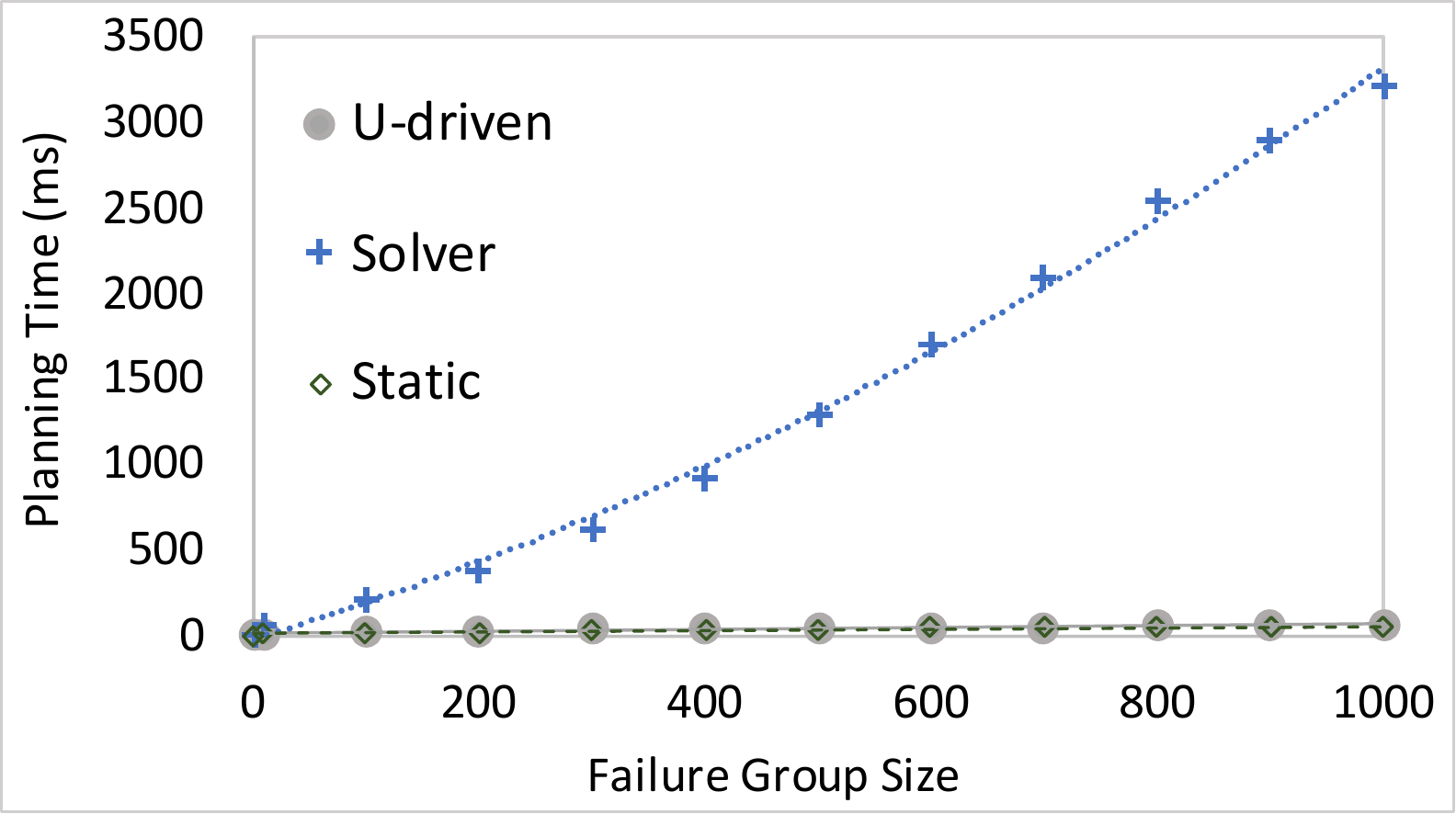}
  \captionof{figure}{Planning time of the three approaches.}
  \label{fig:timran}
\end{minipage}%
\begin{minipage}[t]{.5\textwidth}
  \centering
  \includegraphics[width=0.99\linewidth]{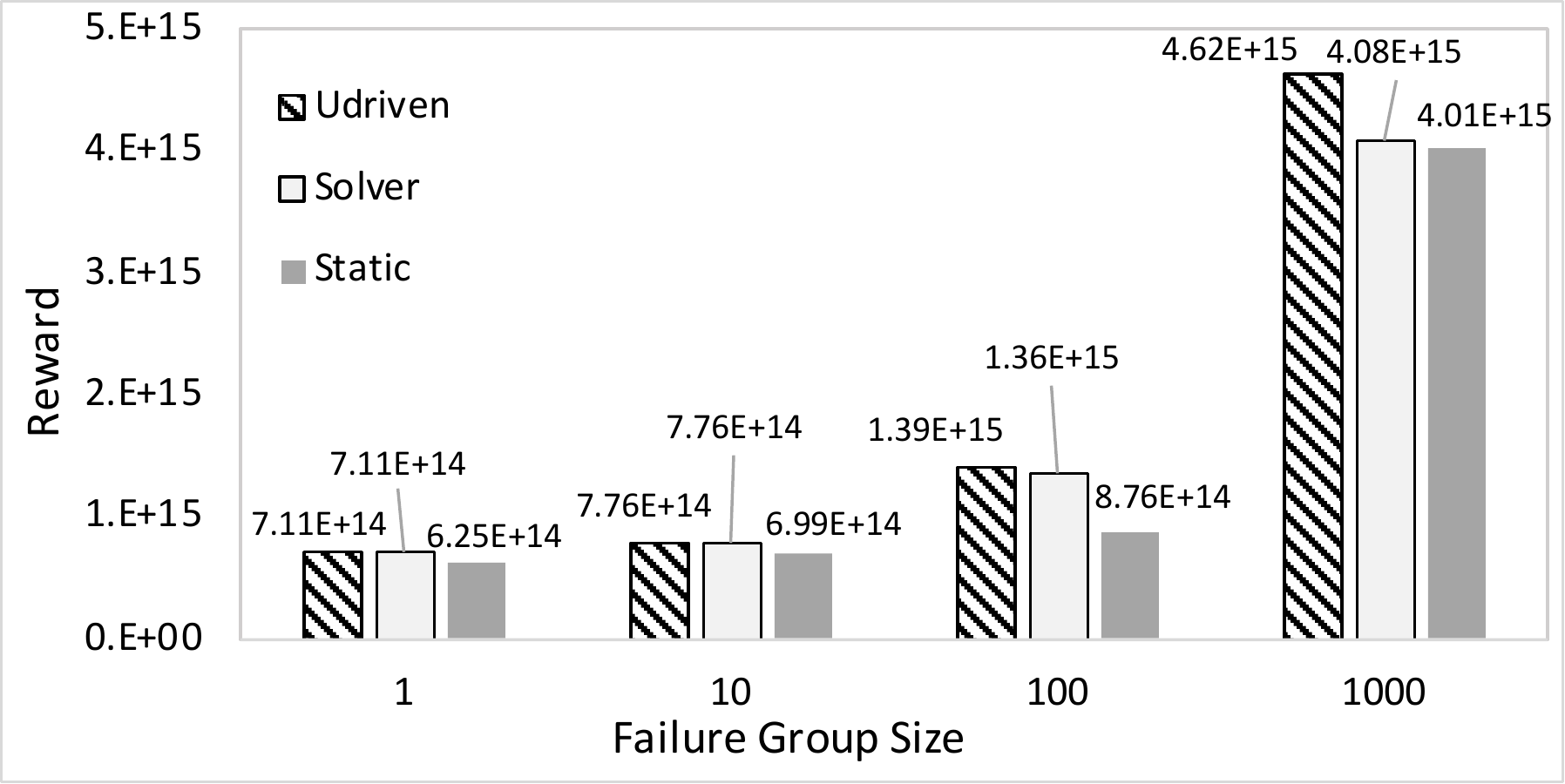}
  \captionof{figure}{Reward of the three approaches over 50 \mbox{MAPE-K} runs.}
 \label{fig:randomrew50}
\end{minipage}
\end{figure}

As the number of injected failures increases, the planning time of all approaches typically increases as well. However, this growth is more extreme for the solver that has to solve larger optimization problems. 
For all of the combinations of numbers of failures and architecture sizes in Table~\ref{fig:Table}, we consider the planning time of the static approach as the \emph{baseline}. 
The planning time of the solver is at least $447\%$ ($10$ failures, $18$ components) and at most $5\text{,}802\%$ ($1\text{,}000$ failures, $1\text{,}800$ components) larger than the \emph{baseline}.
For the u-driven approach, the minimum difference compared to the \emph{baseline} is $9\%$ ($1$ failure, $18\text{,}000$ components) and the maximum is $92\%$ ($100$ failures, $1\text{,}800$ components).
The solver approach always reaches the same optimal configuration as our u-driven approach, but it has an extreme planning overhead in terms of execution time of the planning phase for large numbers of failures and large architectures.
Figure~\ref{fig:timran} visualizes the planning time of the self-healing approaches for the synthetic failure traces. 
To achieve a more precise interpolation, we added additional data points to the ones in Table~\ref{fig:Table}. For this purpose, we extracted additional synthetic failure traces with an FGS of $200$, $300$, .\,.\,. , $900$ from the synthetic failure profile model and conducted further experiments for these traces.
The results show that the solver approach does not scale in contrast to the static and u-driven approaches.

\subsubsection{Experiments for Reward.}\label{subsubsec:performanceSimple}

To compare the reward---that is, the utility over time of the self-healing approaches---we tested them on \mRUBiS with $1\text{,}800$ components. We conducted the experiments for reward on multiple MAPE-K runs~($50$) to make these experiment consistent with the other reward experiments in Sections~\ref{subsubsec:performance-multi} and~\ref{subsubsec:performance-A2}. 
Since we are looking into longer simulation runs ($50$ MAPE-K runs), we observe rather large numbers for the reward. 
The reward of each approach is measured for each of the four synthetic failure traces with an FGS of either $1$, $10$, $100$, or $1\text{,}000$ (cf. Section~\ref{subsubsec:simpleSC}). 
In our earlier work~\cite{Ghahremani+17}, we report on the reward of the self-healing approaches for synthetic failure traces and for single MAPE-K runs. We now extend the single MAPE-K run to map on multiple MAPE-K runs. For this purpose, we consider a duration of 24 hours during which $50$ groups of failures occur while the $50$ MAPE-K runs are uniformly distributed over 24 hours. This results in an IAT of $1\text{,}728$ seconds ($24$ hours divided by $50$). We manually confirmed that $1\text{,}728$ seconds is long enough for the IAT for the most expensive case---that is, the solver approach for the failure trace with an FGS of $1\text{,}000$ and thus with the largest failure density.

Figure~\ref{fig:randomrew50} depicts the results of the experiments in terms of reward of the three self-healing approaches for each of the four failure traces. The results refer to 50 MAPE-K runs, and the IAT is $1\text{,}728$ seconds. 
As the FGS increases, the difference between the reward of the u-driven and solver approaches becomes larger. This is due to planning overhead of the solver as discussed in the context of Figure~\ref{fig:analyticsol}. Larger numbers of failures cause more overhead and therefore a larger loss of reward for the solver approach. This effect is not visible in traces with a smaller FGS of $1$ and $10$.
For all traces and FGS, the reward of the static approach is less than the reward of the u-driven approach. Non-optimal decisions and wrong ordering of the adaptation rules in the static approach cause the loss of reward as discussed earlier with Figures~\ref{fig:analyticstatic} and~\ref{fig:analyticsol}. The loss of reward due to non-optimal decisions can be very severe for a large IAT since the system is performing with the non-optimal utility for a considerable long time.
Similarly, the solver approach obtains for all traces a higher reward than the static approach because it always has enough time (i.e., the IAT is large enough) to optimally resolve the current failures within one feedback loop run before the next group of failures occurs.

\subsection{Experiments for Multiple MAPE-K Runs}\label{subsec:exteval}

The experiments in Sections~\ref{subsubsec:Eresult} and~\ref{subsubsec:performanceSimple} reflect on the scalability and reward of the self-healing approaches for the failure traces extracted from a \emph{synthetic} failure profile model. Although these traces are useful for a preliminary analysis of the scalability and reward, the results of the evaluation are difficult to generalize because they are only based on a single failure profile model. In addition, this failure profile model is \emph{synthetic} and not based on any real-world data.

As described in Section~\ref{subsec:single}, an assumption of the \emph{synthetic} failure profile model is that IAT is large enough so that all occurring failures can be resolved by a MAPE-K run before a new group of failures will occur. However, there is no guarantee that this assumption always holds, and it is a simplification of realistic failure traces. To have more robust and generalizable findings and to explore more diverse characteristics of failure profile models, we use several \emph{realistic} failure profile models with various IAT values to generate \emph{realistic} failure traces for the experiments in this section. 

In the following, we first elaborate on the importance of certain characteristics of failure profile models for the reward using an analytical experiment. We then introduce several \emph{realistic} failure profile models that we use for the experiments and describe how we generate \emph{realistic} failure traces from them. Finally, we investigate how the scalability and reward of different self-healing approaches are influenced by different realistic failure profile models.

\subsubsection{Analytical Experiment.}\label{subsubsec:Analytic-Multi}

This analytical experiment investigates the impact of characteristics of failure profile models for the reward. The experiment uses \mRUBiS with 100 shops (1,800 components) and randomly injects multiple failures of type \elem{CF1}, \elem{CF2}, and \elem{CF3} as a failure group causing a drop of the utility of \mRUBiS. The utility drop is followed by one MAPE-K execution. This MAPE-K run plans for and resolves all of the existing failures. In the case that new failures are occurring while the MAPE-K cycle is running, the planner will not take these failures into account. Nevertheless, these newly occurring failures still cause a utility drop even if they are not yet considered by the planner.

\begin{figure}[t]
	\begin{centering}
		\includegraphics[width=0.75\linewidth]{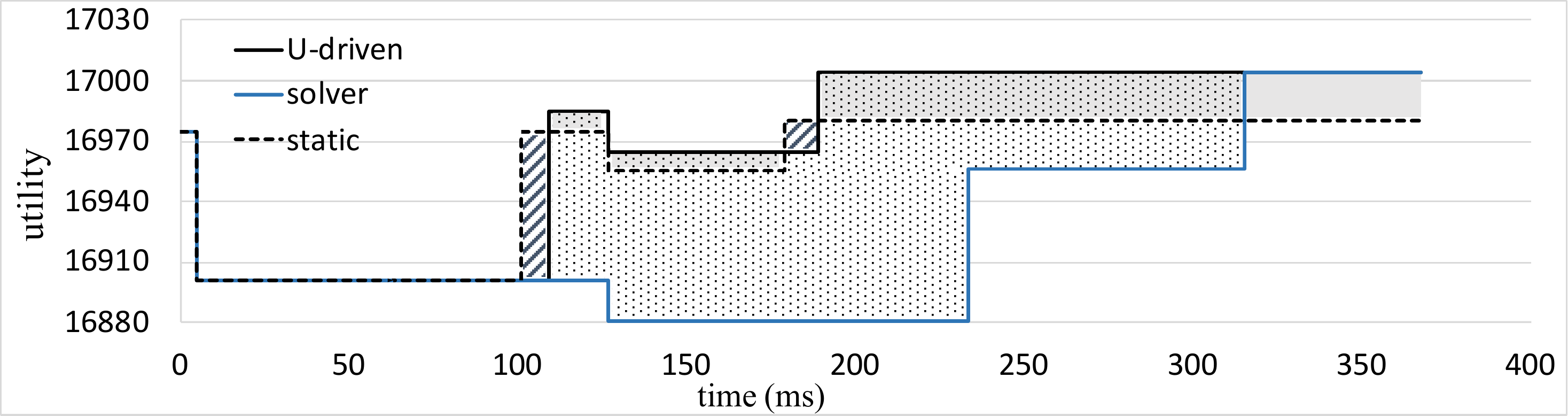}
		\caption{Lost reward of the solver approach compared to the u-driven approach due to longer planning time for a short IAT.}
		\label{fig:an3}
	\end{centering}
\end{figure}

Figure~\ref{fig:an3} shows a variant of the experiment presented in Figure~\ref{fig:analyticsol} regarding the loss of reward of the solver approach compared to the u-driven approach because of the solver's overhead in planning time. In this experiment, this overhead is so large that new failures occur before the solver finishes the current planning. Thus, IAT is shorter than the time required by the solver approach to resolve all existing failures.
Considering Figure~\ref{fig:an3}, the first drop in utility is followed by the first MAPE-K execution where all of the approaches plan for repairing the detected failures.   
Both the static and the u-driven approaches are fast enough to resolve the failures before the next group of failures occurs. Due to the longer planning time of the solver, the solver approach misses the on-time detection of the second group of failures. Therefore, the utility drop caused by the new group of failures remains in the system until they are detected and resolved during the second, delayed MAPE-K run. In contrast, the static and u-driven approaches have already resolved the first group of failures and obtained the increase in the utility by the time the second group of failures occur.
The dotted areas in Figure~\ref{fig:an3} represent the lost reward of the solver approach compared to the u-driven approach.
The gray regions represent the lost utility of the static approach compared to the u-driven approach, whereas the hachured regions depict the gained utility of the static approach compared to the u-driven approach.
This experiment clarifies that in situations where IAT is shorter than the repair time, there will be an additional loss of reward if the planning or the subsequent execution phase overlaps in time with the occurrences of failures.
This is because newly occurring failures in the system are temporarily neglected and hence make the system perform with a lower utility.

\subsubsection{Experiment Design.}\label{subsubsec:multi-scenario}

As revealed by the foundational work on characterizing failure profile models in computer systems~\cite{1676063,192214,1676070}, failures in such systems often have a \textit{bursty} characteristic due to the effects of failure propagation. A single failure in the system triggers a sequence of failures in other system components within a short period of time.
Numerous fault-tolerant algorithms make the assumption that failures occur independently~\cite{Heath:2002:ICA:511399.511362,Zhang2005}. 
This assumption contradicts bursty failure profile models and neglects that occurrences of failure bursts often result in correlated availability behaviors of different components.
\citeN{Iosup:2006:RGU:1513923.1513971} showed that ignoring the bursty character of failure profile models results in overestimating the transient reward rate by an order of magnitude, even if only as few as $10\%$ of the failures conform to a bursty profile~model.

In the following, we describe \emph{realistic} failure profile models that are based on failure traces of real-world computer systems and that we will use for the evaluation.
Table~\ref{tab3} compares the characteristics of the \emph{synthetic} failure profile model used in Section~\ref{subsec:single} to the \emph{realistic} models used in this section.
FGS and IAT of the failure traces derived from the realistic failure profile models vary for each MAPE-K run while they are constant and large enough in the synthetic model.
The FET is larger than $0$ in the realistic failure profile models so that failures take some time to propagate in the system. In contrast, the FET is $0$ in the synthetic failure profile model so that all failures of a group occur at once.
 
\begin{table}[t]\centering
	\ra{.9}
	\caption {General Characteristics of Different Failure Profile Models}
	\label{tab3}
	\vspace{-1em}
	\scalebox{.9}{
	\begin{tabular}{lcc}\toprule
		& \multicolumn{1}{c}{\textbf{Synthetic Failure Profile Model}} & \multicolumn{1}{c}{\textbf{Realistic Failure Profile Models}} \\
		\cmidrule{2-3}	
		
		Failure Group Size  (FGS) & Constant  & Varies for each MAPE-K run \\
		
		Inter Arrival Time (IAT) & Large enough   & Varies for each MAPE-K run \\
		
		Failure Exposure Time (FET) &$0$& Larger than $0$\\
		
		\bottomrule
	\end{tabular}
	}
\end{table}

\paragraph{Realistic failure profile models.}

To investigate the impact of realistic failure profile models on the scalability and reward of the self-healing approaches, we studied three different models. These models are constructed from real-world failure traces provided by \citeN{Gallet2010}, originate from different computer systems, and differ in scale and volatility. As mentioned previously, failure traces are derived from failure profile models for a certain duration, and the failure density of a trace refers to the overall number of failures within this duration.

Of all failure profiles models~\cite{Gallet2010}, we selected three models with different failure densities and sizes of the originating system: \grid, \lri, and \deug.
The \grid model originates from the \grid system, in which a significant fraction of failures occurs in bursts. \grid is an experimental grid environment with more than $2\text{,}500$ processors and $1\text{,}288$ nodes \cite{iosup:inria-00143265}, which is comparable to the size of \mRUBiS with 100 shops (1,800 components). The event data for \grid has been gathered over $1.5$ years of monitoring~\cite{Kondo:2010:FTA:1844765.1845157}. 
The other models, \deug and \lri, are constructed from application-level traces of real enterprise desktop grids that contain bursts of failures~\cite{Gallet2010} and have been collected over $1$ month from about $100$ (\deug) and $40$ (\lri) hosts~\cite{kondo_fgcs07}.

\begin{figure}[t]
	\begin{centering}
		\includegraphics[width=0.36\linewidth]{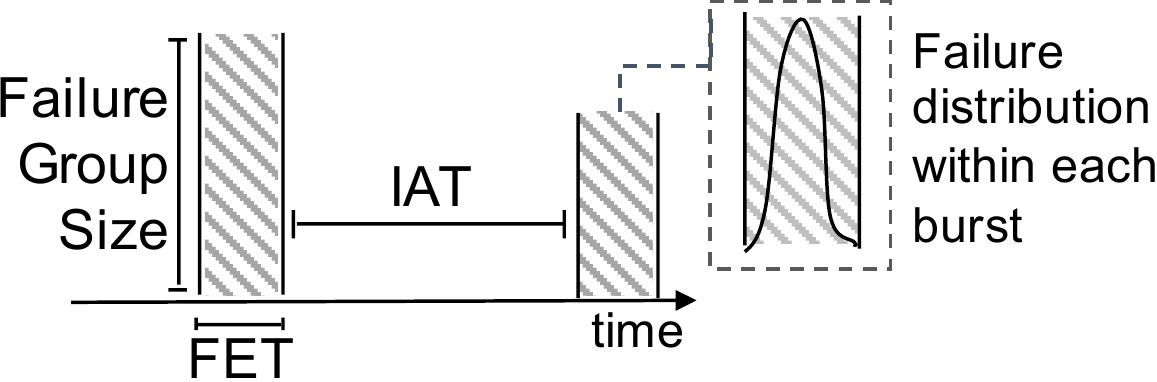}
		\vspace{-1em}
		\caption{Failure group size (FGS), failure exposure time (FET), and inter arrival time (IAT) of failure profile models.}
		\label{fig:burst}
	\end{centering}
\vspace{-1em}
\end{figure}

All three failure profile models fit statistical distributions to IAT and FGS. The latter is the failure group size that corresponds to the number of failures that occur in a burst---that is, in a group of failures. Moreover, \citeN{Gallet2010} consider different window sizes for monitoring and detecting the failure occurrences in each model. This window size is equivalent to the failure exposure time (FET) illustrated in Figure~\ref{fig:burst}. Thus, each burst occurs within the FET. However,  \citeN{Gallet2010} do not clarify how the failures are distributed within each burst. Therefore, we assume that failures propagate following a normal distribution during each burst (cf.~Figure~\ref{fig:burst}). 
Table~\ref{tab2} lists the distributions proposed by \citeN{Gallet2010} for IAT and FGS for the three failure profile models along with the considered FET.

\begin{table}[t]\centering
 	\ra{1.2}
 	\caption{Characteristics of the Realistic Failure Profile Models \lri, \deug, and \grid and the Variants Uniform, Single, and Bigburst of the \grid Model.}
	\label{tab2}
	\scalebox{.67}{
 	\begin{tabular}{lrrr|rrr}\toprule
 		& \multicolumn{1}{c}{\textbf{LRI}} & \multicolumn{1}{c}{\textbf{DEUG}} & \multicolumn{1}{c|}{\textbf{Grid5000 (Burst)}}
 		& \multicolumn{1}{c}{\textbf{Uniform}} & \multicolumn{1}{c}{\textbf{Single}}& \multicolumn{1}{c}{\textbf{Bigburst}} \\
 		\cmidrule{1-7}
 				
 		FGS & $LOGN(1.32,0.77)$  &$LOGN(2.15,0.70)$&$LOGN(1.88,1.25)$ &$N(22.85,20.68)$&$1$&$N(238,97.3)$\\
 		
 		IAT (seconds) & $LOGN(-1.46,1.28)$ & $LOGN(-2.28,1.35)$&$LOGN(-1.39,1.03)$ &$1728$&$77.4$~&$N(3521.4,5418.6)$\\
 		
 		FET (seconds)&$100$&$150$&$250$&$250$&N/A&$250$\\ \midrule
 		
 		No. of bursts $n$ ~(short trace)&$50$&$50$&$50$ &$50$&$1\text{,}116$&$6$\\
 		
 		No. of bursts $n$ ~(long trace)&$1\text{,}355$&$2\text{,}843$&$1\text{,}678$&--&--&--\\
 		
 		Duration (hours)~(short trace)& $41.2$&$21.4$&$24$&$24$&$24$&$24$\\
 		
 		Duration (days)~(long trace)&30&30&30&--&--&--\\
 		
 		Failure Density $d$~(short trace)&$318$&$666$&$1\text{,}116$&$1\text{,}116$&$1\text{,}116$&$1\text{,}116$\\
 		
 		Failure Density $d$~(long trace)&$7\text{,}568$&$32\text{,}895$&$25\text{,}279$&--&--&--\\
		
 		\bottomrule
 	\end{tabular} 
	}
\end{table}

From each realistic failure profile model \lri, \deug, and \grid, we extracted a short and a long failure trace.
As shown in Table~\ref{tab2}, the short traces include $n=50$ occurrences of bursts. Their durations are different ($41.2$, $21.4$, and $24$ hours) for each model due to different distributions of IAT. The long traces last for $30$ days and differ in the number of bursts ($1\text{,}355$, $2\text{,}843$, and $1\text{,}678$) for each model. Finally, Table~\ref{tab2} lists the failure densities of the traces.

\paragraph{Variations of realistic failure profile models.}

To ensure a fair and meaningful comparison between the experiment results for the different models, we extracted traces from these models that have the same failure density of $d=1\text{,}116$ (as in the short trace of the original \grid model in Table~\ref{tab2}). We modified parameters of the \grid failure profile model (cf.~Table~\ref{tab2}) to obtain more variants with extreme characteristics. Using these variants, we study the impact of extreme characteristics of failure profile models on the scalability and reward of different self-healing approaches. In addition, using more and extreme failure traces for the experiments allows us to evaluate the robustness of the approaches and results. 
Based on the original \grid model that is named the \emph{burst} model (cf.~\citeN{Gallet2010}), we constructed three modified failure profile models listed in Table~\ref{tab2}:
(1)~the \emph{uniform} model in which failures are uniformly distributed, 
(2)~the \emph{single} model with a single failure at each burst, 
and (3)~a \emph{bigburst} model with only large bursts. 

\paragraph{Burst model.}

\begin{figure}[t]
	\begin{centering}
		\includegraphics[width=0.61\linewidth]{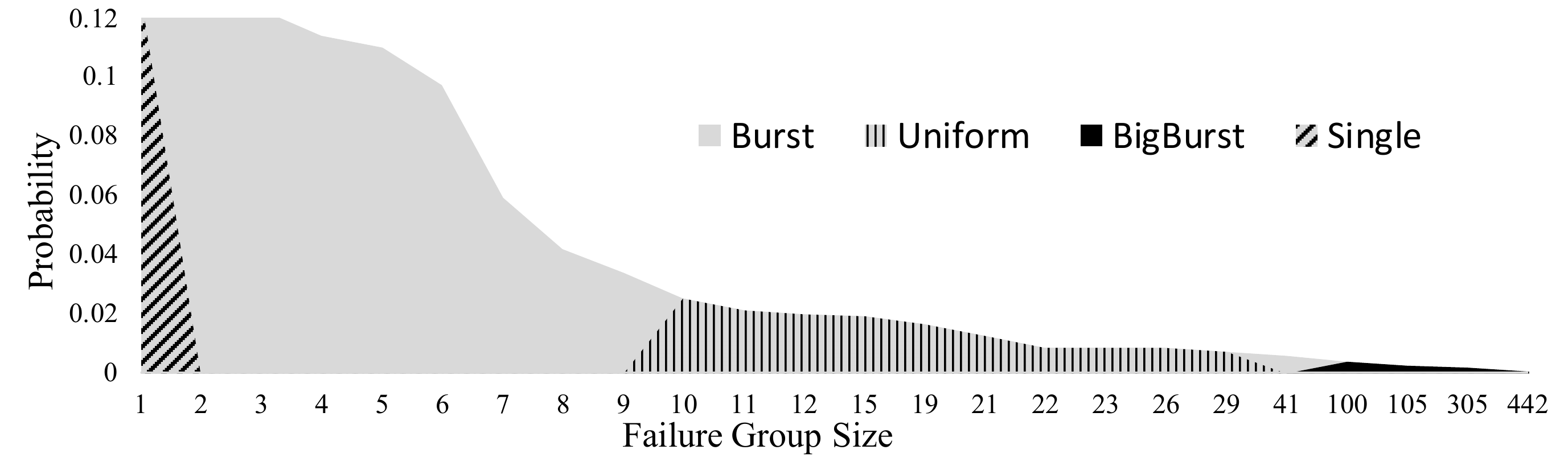}
		\vspace{-1em}
		\caption{Failure group size (FGS) distributions for the variants of the \grid failure profile model.}
		\label{fig:dist}
	\end{centering}
\end{figure}

For the original \grid failure profile model as provided by \citeN{Gallet2010}, we generate the same \emph{short} trace as shown in Table~\ref{tab2} with $n=50$ occurrences of failure bursts. Given the statistical distribution of IAT, it takes $24$ hours for the $50$ failure bursts.
Figure~\ref{fig:dist} shows the generated FGS distribution for the burst model.

\paragraph{Uniform model.}\label{subsubsec:uniform-failure}

We construct a uniform failure profile model from the \grid model, which has the same failure density $d$ during $24$ hours as the burst model, and which has a uniform distribution for IAT and FGS.
To construct this model, we consider the original set of $50$ failure groups extracted from the \grid model for the short trace (see Table~\ref{tab2}) as the set $S$.
A normal sample distribution is extracted from $S$ using statistical bootstrapping~\cite{Davison2002,efron93bootstrap}.
For this purpose, we randomly re-sampled $S$ using statistical bootstrapping and formed a new set $S'$ of the mean values of each sample set. A normal distribution $N(\mu_{S'}, 2\sigma^{2}_{S'})$ is used to generate random values for FGS. 
The resulting set consists of uniformly distributed values within a certain margin extracted from the original set $S$. Applying this distribution, we generated a sequence of normally distributed values for FGS while keeping the same number of bursts $n$ as in the burst model. Thus, the uniform model takes all $d$ failures within $24$ hours and distributes them among $n$ occurrences by using the extracted uniform distribution. Figure~\ref{fig:dist} sketches the range of the original FGS (in the burst model) that is also present in the uniform model.
The IAT is the average of the IAT values in the burst model. Therefore, the uniform model is a sequence of failure groups with normally distributed sizes that occur in equal intervals. As mentioned earlier, the failure density $d$ is the same as in the burst model (see Table~\ref{tab2}). 

\paragraph{Single model.}\label{subsubsec:single-failure}

To consider a naive failure profile model that---to the best of our knowledge---has been used in many existing work on self-healing (e.g.,~\citeN{Carzaniga:2008:SMA:1370018.1370023,Casanova:2013:DAR:2487336.2487354,Angelopoulos:2014:DMF:2593929.2593936,Magalhaes:2015:SWS:2744297.2700325,Perino:2013:FSS:2486788.2487016,DiMarco:2013:SSC:2487336.2487358}), we construct the single failure profile model. In this model, failures are not correlated so that they arrive individually and not in groups. Thus, $FGS = 1$ (cf.~Figure~\ref{fig:dist}). In the failure trace extracted from this model, individual failures are equally distributed within the $24$ hours keeping the same failure density $d$. The number of occurrences of bursts $n$ equals $d$ since each burst includes exactly \emph{one}~failure (see Table~\ref{tab2}).

\paragraph{Bigburst model.}\label{subsubsec:only-burst-failure}

We also consider the other end of the spectrum---that is, occurrences of large failure bursts with $150$ to $450$ failures at each burst. Similar to the construction of the uniform model, we use statistical bootstrapping to extract a corresponding set from the original set $S$. To achieve large FGS, only the part of $S$ that is above a certain threshold (i.e., FGS $\geq$ 100) has been re-sampled for bootstrapping (see Figure~\ref{fig:dist}). To keep the same failure density $d$, the number of bursts decreases to $n=6$ since the failures occur only in large group sizes. Hence, IAT increases accordingly. 
The IAT values for this model are extracted by bootstrapping from the randomly re-sampled IAT values where IAT $\geq 1\text{,}000$ seconds from the IAT values of the original \grid model (see Table~\ref{tab2}).

\subsubsection{Experiments for Scalability.}
\label{subsubsection:scalability-multi}

\begin{figure}[t]
	\centering
	\begin{minipage}[t]{.42\textwidth}
		\centering
		\includegraphics[width=1\linewidth]{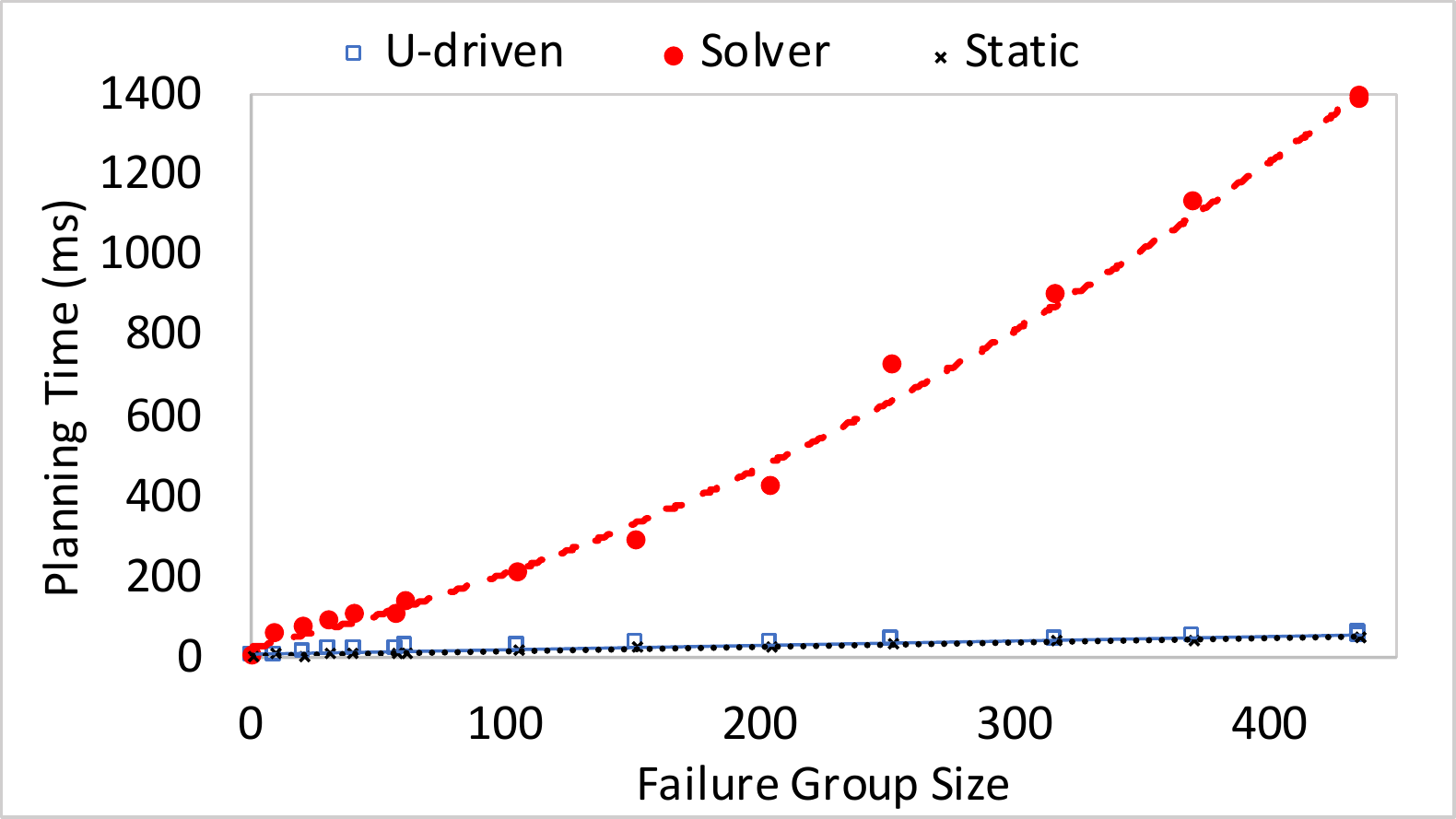}
		\captionof{figure}{Planning time for all failure profile models.}
		\label{fig:timrest}
	\end{minipage}%
	\hspace{2em}
	\begin{minipage}[t]{.5\textwidth}
		\centering
		\includegraphics[width=1\linewidth]{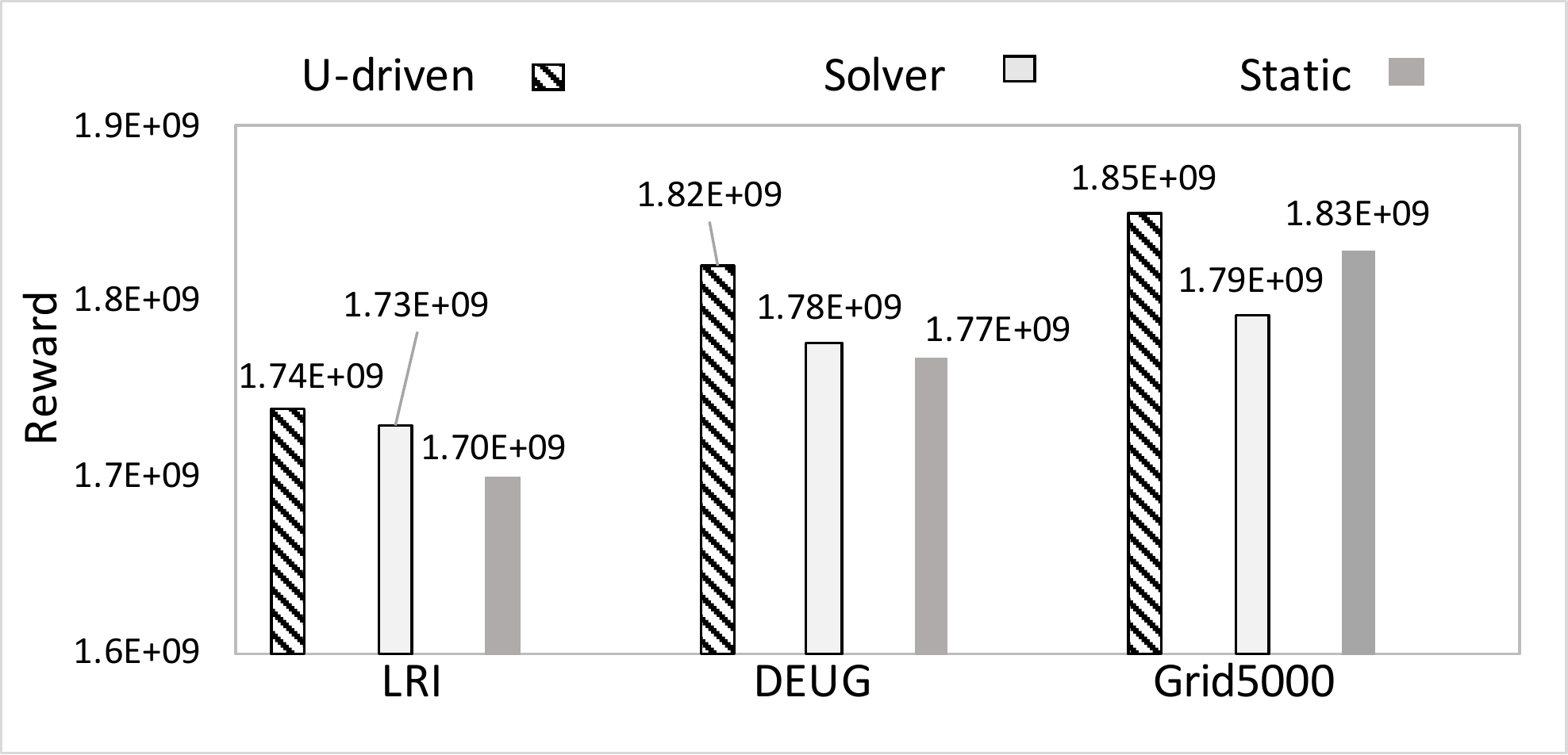}
		\captionof{figure}{Reward for \textit{short} traces of realistic failure profile models.}
		\label{fig:allB}
	\end{minipage}
\vspace{-1em}
\end{figure}

In the following, we investigate whether using the traces extracted from the \textit{realistic} failure profile models and their modified variants (see~Table~\ref{tab2}) confirm the scalability results that we obtained for the \textit{synthetic} models in Section~\ref{subsubsec:Eresult}. We used \mRUBiS with $100$ shops ($1\text{,}800$ components) and conduct the same scalability experiments as before. The results are as follows.

Figure~\ref{fig:timrest} shows the planning time for each of the three self-healing approaches considering all of the six short failure traces, each generated from the different failure profile models listed in Table~\ref{tab2}. These results are the averages of the planning time in milliseconds over $300$ repetitions for each of the six traces.
Employing failure traces of the single or uniform model results in a large population of data points~(i.e., planning time measurements for certain FGS values) for $\text{FGS}<50$ (see FGS distributions for the uniform and single models in Figure~\ref{fig:dist}).
Therefore, in Figure~\ref{fig:timrest}, to avoid optimizing the interpolation curve for the range of $1\leq\text{FGS}<50$, we randomly sampled for all three self-healing approaches the data points for $\text{FGS}<50$. 
Consistent with the results for the synthetic failure profile models (cf.~Section~\ref{subsubsec:Eresult}), the u-driven approach has a lower overhead in terms of planning time than the solver approach, and it is close to the static approach that does not require much runtime planning effort. This also holds for large numbers of failures. Similar to Figure~\ref{fig:timran}, both the static and u-driven approaches have linear growth in planning time as the FGS increases. However, the planning time of the solver approach increases with a polynomial gradient as the FGS increases. Therefore, we can confirm the tendency observed for the synthetic failure traces in Figure~\ref{fig:timran} by the results shown for the realistic failure traces in Figure~\ref{fig:timrest}: the solver approach does not scale well in contrast to the static and u-driven approaches.

\subsubsection{Experiments for Reward.}
\label{subsubsec:performance-multi}

The following experiments compare the reward of the three different self-healing approaches using the realistic failure profile models and \mRUBiS with $100$ shops~($1\text{,}800$ components). 
Each simulation is conducted for the short~($n$ = $50$ occurrences of bursts lasting roughly between $21$ and $41$ hours) and long~($30$ days with different but fixed numbers of occurrences) traces generated from the \lri, \deug, and \grid failure profile models, as well as the short traces generated from the uniform, single, and bigburst models of \grid (cf.~Section~\ref{subsubsec:multi-scenario} and Table~\ref{tab2}).

\begin{figure}[t]
	\centering
	\begin{minipage}{.45\textwidth}
		\centering
		\includegraphics[width=1\linewidth]{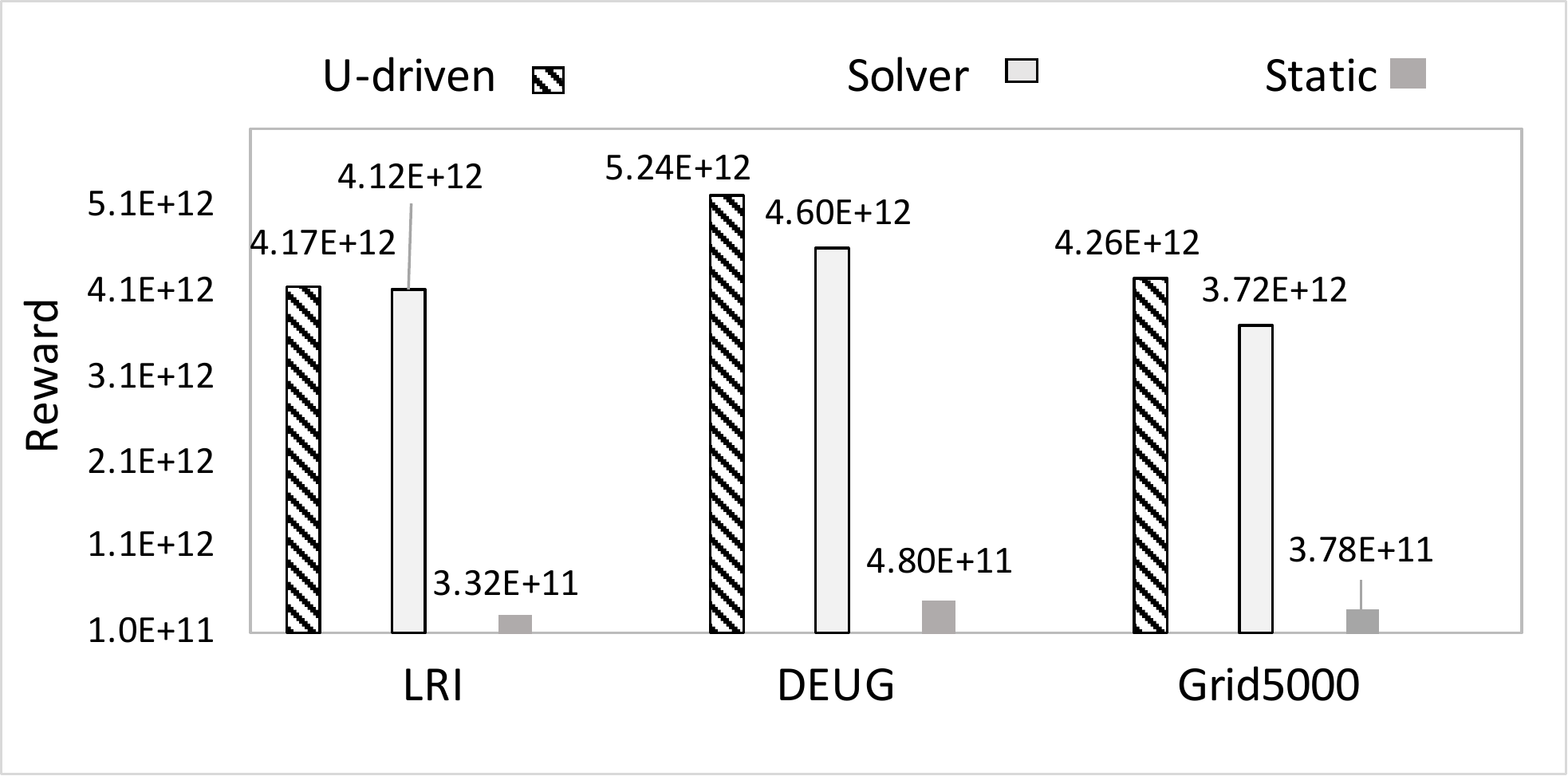}
		\captionof{figure}{Reward for \textit{long} traces of realistic failure profile models.}
		\label{fig:rew30days}
	\end{minipage}%
	\hspace{0.5em}
	\begin{minipage}{.49\textwidth}
		\centering
		\includegraphics[width=.93\linewidth]{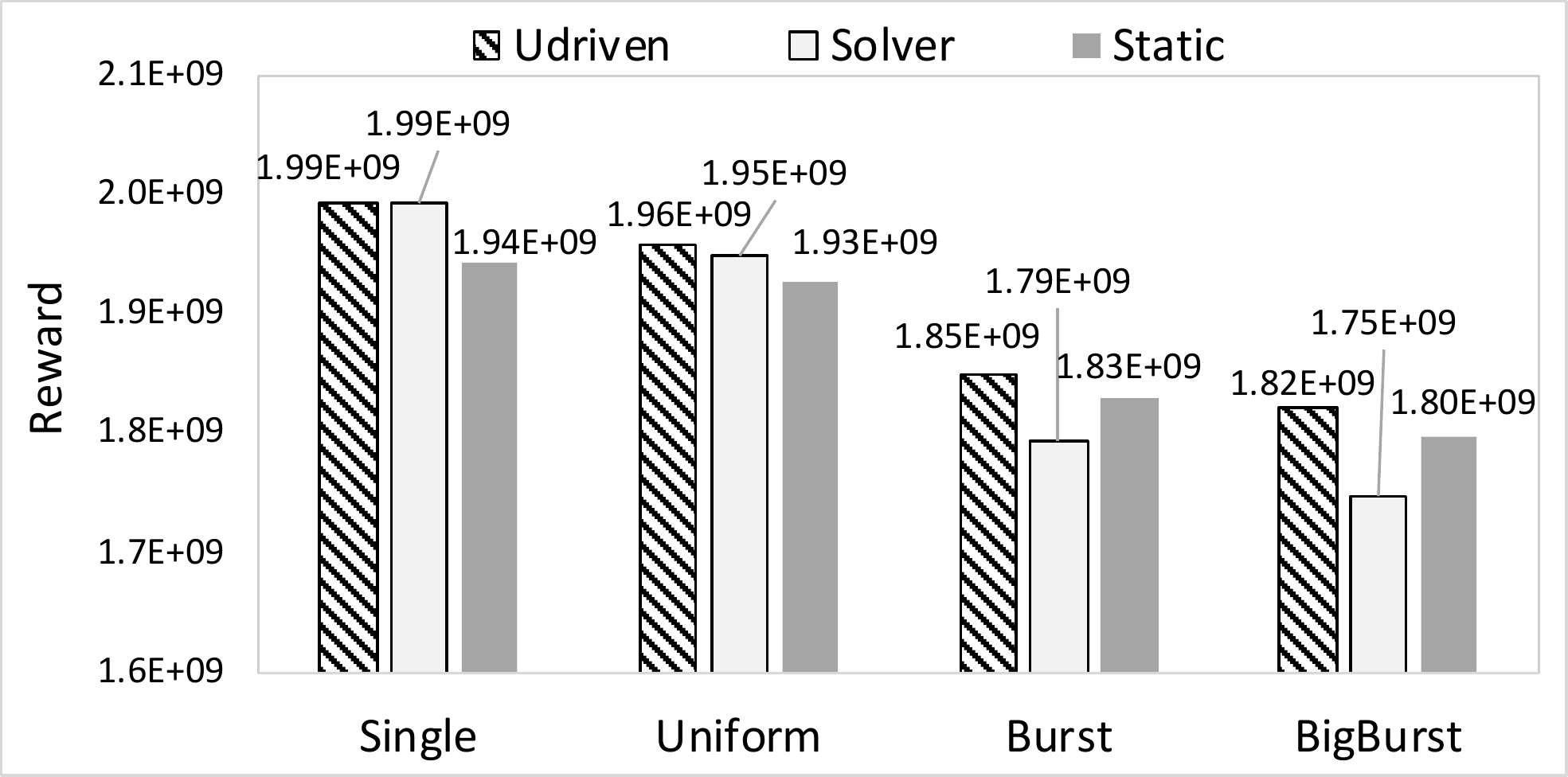}
		\caption{Reward for traces of the modified \grid model.}
		\label{fig:FPM2}
	\end{minipage}
\end{figure}

Figure~\ref{fig:allB} shows the reward of the self-healing approaches for the \textit{short} traces of the \lri, \deug, and \grid models.
Since these models have different characteristics (e.g., the generated traces have different failure densities for each model), the results cannot be compared across them. However, we can compare the results in terms of the reward achieved by the three self-healing approaches within each model and trace.
As shown previously, if many failures occur, the solver approach considerably requires more time for planning than the u-driven and static approaches.
Consequently, for the short \grid trace with a total of $1\text{,}116$ failures, the solver approach achieves the lowest reward compared to the other two approaches (see Figure~\ref{fig:allB}). 
In contrast, for the short \deug and \lri traces with a total of $666$ and $318$ failures, the solver approach performs slightly better than the static approach.
Thus, the different failure densities of the traces (cf. Table~\ref{tab2}) influence the performance of the different self-healing approaches.

Moreover, if the IAT is shorter than the time an approach needs to resolve the failures, there will be a more severe loss of reward (cf.~Figure~\ref{fig:an3}). This effect applies to the solver approach because of the costly planning, which is a further reason that this approach achieves a considerably lower reward than the other approaches for the \grid trace with a high failure density while not to such an extent for \lri and \deug traces with lower failure densities.
For the \lri model that has the lowest failure density, the performance of the solver approach is close to the u-driven approach, whereas this gap is larger for the \deug and \grid traces that have higher failure densities (see Figure~\ref{fig:allB}).

Figure~\ref{fig:rew30days} shows the reward of the self-healing approaches for the \textit{long} traces generated from the \lri, \deug, and \grid models. 
Using these traces covering roughly a period of $30$ days, we can observe the longer execution of the approaches. Particularly, we observe that the static approach loses severely more reward than the other two approaches for all failure profile models. 
As discussed for the analytical experiments in Section~\ref{subsubsection:Analytic-single}, the impact of non-optimal decisions made by the static approach on the reward can remain permanently in the system. Therefore, the reward loss due to non-optimal decisions remains and propagates through $30$ days of execution, which considerably reduces the achieved reward.
Similar to the results in Figure~\ref{fig:allB}, the solver approach achieves less reward than the u-driven approach due to the planning overhead. However, the difference is that the reward achieved by the solver approach is considerably larger than the reward achieved by the static approach. The reward loss due to the overhead of the solver approach seems to be compensated over time and does not severely impact the system.

As the failure densities of the traces differ (cf.~Table~\ref{tab2}), the reward achieved by the self-healing approaches for one trace cannot be compared to the reward for a different trace. To enable a comparison across traces, we use traces with an equal failure density. Therefore, we use the modified variants of the \grid model to generate traces with the same failure density (cf.~Section~\ref{subsubsec:multi-scenario}). 
The reward achieved by the self-healing approaches for the modified variants  (single, uniform, and bigburst) and the original \grid model (burst) is presented in Figure~\ref{fig:FPM2}. 
As discussed for the analytical experiment in Section~\ref{subsubsec:Analytic-Multi} and confirmed by these results, certain characteristics of the failure profile models influence the reward of the self-healing approaches.
The solver approach achieves the lowest reward among all of the three approaches for the burst and bigburst models whose traces have a large FGS.
Although the difference between the reward of the u-driven and static approaches is small, the u-driven is still achieving a larger reward for these two models.
For the bigburst model, the overhead of the u-driven approach causes the reward loss compared to the burst model. 
The difference between the reward of the solver and static approaches is also larger for bigburst than for the burst model. The performance of the solver approach is negatively dominated by the planning overhead. The static approach achieves a larger reward than the solver approach, although it is not optimal, as confirmed in Section~\ref{subsec:single}.

For the uniform model, the solver approach achieves more reward than the static but less than the u-driven approach. In this case, the FGS is smaller than in the burst and bigburst models so that the impact of the solver's costly planning is less severe. However, there is still some loss of reward compared to the u-driven approach because of the lower overhead of the u-driven approach.
The single model does not affect the reward of the solver and u-driven approaches because it is a simple model (i.e., at each arrival of failures, there is only a single failure to repair). In this case, the static approach achieves the lowest reward among the three approaches (cf.~Figure~\ref{fig:FPM2}) since it is not optimal in selecting the best adaptation rule for the failure (this corresponds to the impact of non-optimal decisions shown in Figure~\ref{fig:analyticstatic}(a)).

Summing up, the experiments with the \textit{realistic} failure profile models having different characteristics regarding FGS, IAT, FET, failure density, and duration (cf.~Table~\ref{tab2}) show that the static and u-driven approaches scale well in contrast to the solver approach and that the u-driven approach outperforms---in terms of achieved reward---the static approach in general and the solver approaches in cases where failures occur in bursts. Only in the single model do the u-driven and solver approaches achieve the same reward (cf. Figure~\ref{fig:FPM2}), because no costly planning is required when there is only a single issue to resolve. These results confirm the results obtained for the \textit{synthetic} failure profile model in Section~\ref{subsec:single}.

\subsection{Possible Violation of Assumptions}\label{subsec:violation-of-assumption-A2}

In this work, we made several assumptions, as listed in Table~\ref{tab1}. We discussed in Section~\ref{subsec:discussions-for-assumptions} that these assumptions are usually justified for rule-based self-healing approaches.
For instance, we assume a deterministic and effective behavior of adaptation rules that is always able to repair the occurred failures (assumption (A2)). However, there might be cases in which rules will not always succeed in repairing failures, which violates assumption (A2).  
We will therefore investigate such cases by considering probabilistic adaptation rules---that is, each rule has a likelihood for its success in resolving a failure.
If a rule is not successful, the failure remains in the system and is dealt with during the next MAPE-K run.

\subsubsection{Analytical Experiments.}\label{subsubsec:Analytic-A2}

For the analytical experiments, we use \mRUBiS with 100 shops (1,800 components). The experiment starts with random occurrences of multiple failures of type \elem{CF1}, \elem{CF2}, and \elem{CF3} as one failure group, which causes a drop of the utility of \mRUBiS. The utility drop is followed by one or more MAPE-K runs.

\begin{figure}[t]
	\begin{centering}
		\includegraphics[width=0.99\linewidth]{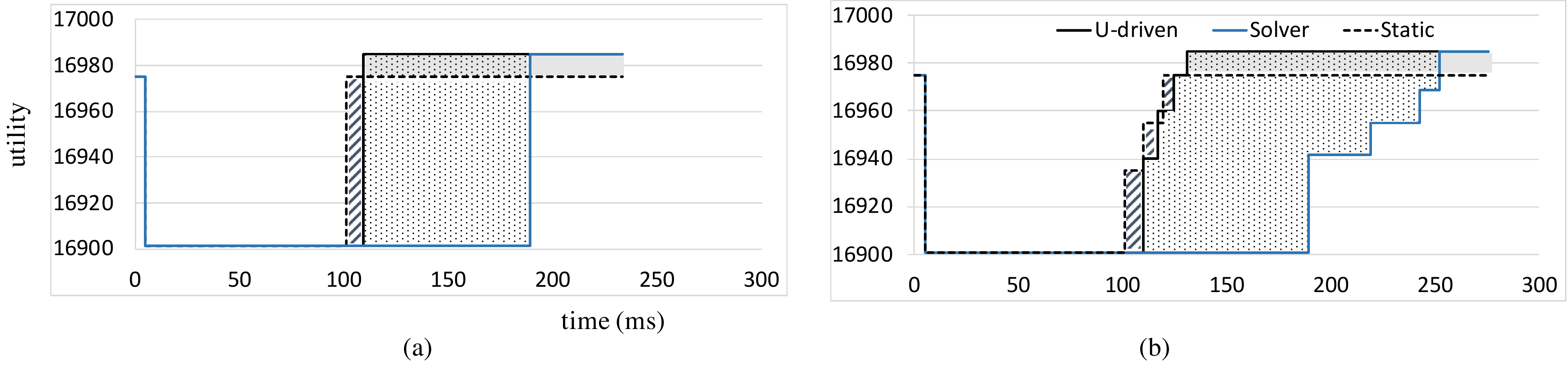}
		\vspace{-1em}
		\caption{Analytical experiment with probabilistic adaptation rules.}
		\label{fig:prob_analytical}
	\end{centering} 
	\vspace{-1em}
\end{figure}

Figure~\ref{fig:prob_analytical}(a) shows the case where the first MAPE-K run plans for and resolves all of the existing failures. All of the applied adaptation rules are $100\%$ effective so that they successfully resolve the failures. Thus, \mRUBiS continues operating with the similar level of utility as before the failures have occurred.
Figure~\ref{fig:prob_analytical}(b) repeats the same experiment, but the success likelihood of all adaptation rules is now set to $50\%$. After the first MAPE-K run, all approaches fail to bring back the utility of \mRUBiS to the level before the failures have occurred. The failures that could not be resolved remain in the system and require additional attempt(s) (MAPE-K runs) to repair them. Each MAPE-K run has one attempt of executing the selected adaptation rules. This delays the point in time until all failures are resolved.
 
Considering Figure~\ref{fig:prob_analytical}(b), the dotted area is the lost reward of the solver approach compared to the u-driven approach. The gray (hachured) areas represent the reward loss (gain) of the static over the u-driven approach. 
In general, the longer a failure remains in the system---for instance, because of ineffective adaptation rules---the larger is the reward loss. 
Moreover, as confirmed in Sections~\ref{subsec:single} and~\ref{subsec:exteval}, the solver approach has a larger planning time and thus requires more time to resolve the failures than the other approaches. Thus, the reward loss is larger for the solver~approach.

\subsubsection{Experiment Design.}\label{subsubsec:A2-SC}

The experiments in this section are designed according to Section~\ref{subsubsec:multi-scenario}, but they are limited to the short trace extracted from the \grid model. We selected this trace since it has the highest failure density among the short traces (cf.~Table~\ref{tab2}). We consider different success likelihoods of either $100\%$, $75\%$, $50\%$, or $25\%$ for all adaptation rules and \mRUBiS with $100$ shops ($1\text{,}800$ components).

\subsubsection{Impact on Scalability.}

Reduced success likelihoods of the adaptation rules require additional MAPE-K runs during the IAT to resolve the remaining failures in \mRUBiS.
Unless IAT is very short such that not enough repair attempts in additional MAPE-K runs can be done, the violation of assumption (A2) will not affect the scalability of the self-healing approaches because the size of the planning problem does not change. 
A violation of assumption (A3a) results in adaptation rules causing new issues (cf. Section~\ref{subsec:discussions-for-assumptions}). This effect introduces additional failures to the originally detected ones. Therefore, it can impact scalability by increasing the size of the failure groups. Such an effect influences all of the self-healing approaches, although not equally. Those with a more costly planning, such as the solver approach, are more likely to suffer from this effect. 
The violation of assumptions (A3b) and (A4) imply the existence of dependencies between the adaptation rules. Detecting and resolving such dependencies can complicate the planning phase and require more exhaustive solutions, such as model checkers, to address this problem. This would impact negatively the scalability of the approach. Since this effect is directly related to the adaptation rules and not to the overall self-healing approaches, the impact will be the same for all approaches particularly if they all use the same technique to resolve these dependencies.
Violation of assumption (A6) can potentially influence the scalability of the proposed scheme, but this will affect \textit{all} self-healing approaches evenly since it will add increased costs equally to all approaches.
 
Finally, a violation of the remaining assumptions (A1) and (A5) will not have any impact on the scalability of the approaches. As discussed in Section~\ref{subsec:discussions-for-assumptions}, the violation of assumption (A1) does not hold for the case of self-healing systems, and the violation of assumption (A5) indicates the need to extend the context for adaptation rules. This happens at design time and does not impact the runtime scalability of the self-healing approaches.  

\subsubsection{Experiments for Reward.}\label{subsubsec:performance-A2}

\begin{figure}[t]
	\begin{centering}
		\includegraphics[width=0.69\linewidth]{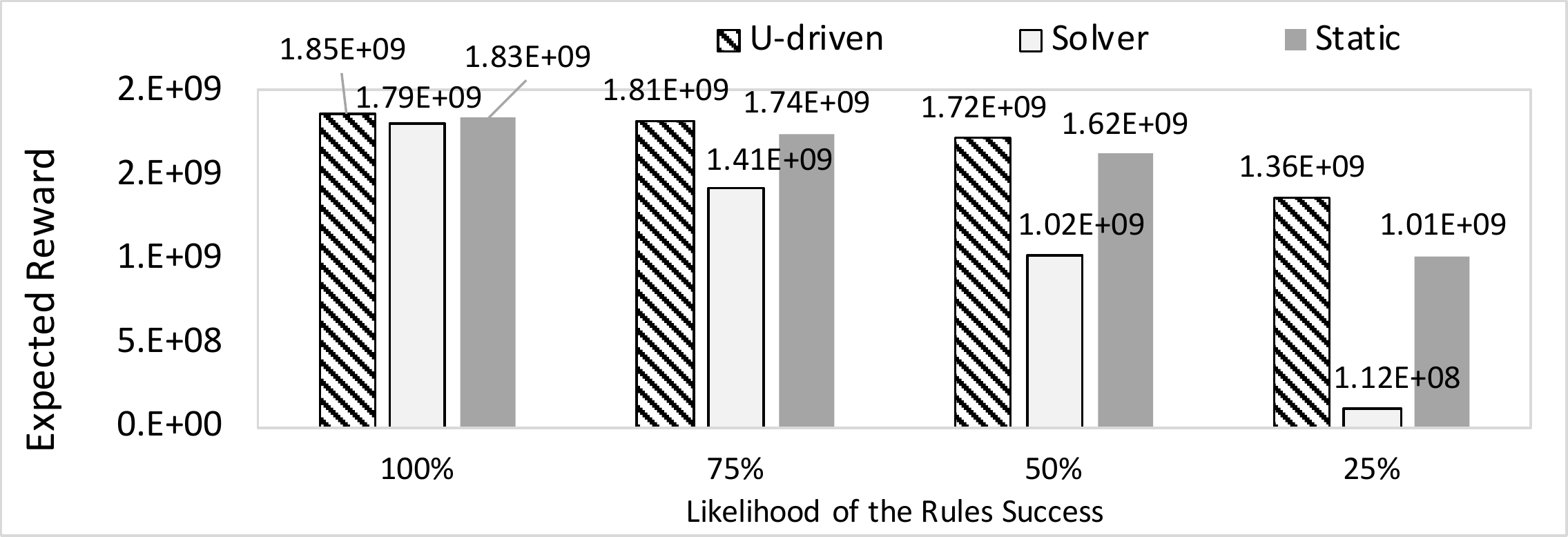}
		\caption{Reward achieved by self-healing approaches with probabilistic rules and different success likelihoods.}
		\label{fig:prob}
	\end{centering} 
\end{figure}

In this section, we demonstrate the impact of violating assumption (A2) on the reward. Figure~\ref{fig:prob} shows the reward achieved by the self-healing approaches with different likelihoods of success for the adaptation rules.
In the case of $100\%$ success likelihood, assumption (A2) holds. As the success likelihood decreases, all self-healing approaches gain a lower reward since failures remain in the system for longer time until they are finally resolved. 
The solver approach, however, loses more reward compared to the static and u-driven approaches because of its costly planning. As the rules fail to resolve the failures, the MAPE-K loop keeps re-planning for the remaining failures with additional runs. Meanwhile, new failures might occur, increasing the number of failures to be repaired and thus the size of the optimization problem to plan the self-healing. Particularly, the performance of the solver approach is mostly affected by the number of failures (cf.~Figures~\ref{fig:timrest} and~\ref{fig:allB}).
Thus, the most severe impact of the success likelihood can be observed for the solver approach due to its costly planning. In cases with small success likelihoods (i.e., $25\%$), more frequent planning is required to resolve the remaining issue and the expected reward of the solver approach drops drastically. The u-driven approach, despite its minor overhead, still manages to outperform the static approach that has no planning overhead. 
Thus, the u-driven approach outperforms the static and solver approaches considering different success likelihoods of adaptation rules and therefore in settings where assumption (A2) is violated.
 
\subsection{Threats to Validity}
\label{subsec:threats-to-validity}

\subsubsection{Internal Validity.}
Threats to internal validity concern how we performed the experiments and interpreted the results. To address such threats, we systematically investigated the scalability and reward of the static, solver, and u-driven self-healing approaches by using the controlled simulation environment \mRUBiS~\cite{2018-mRUBiS} for the experiments. Particularly, we are interested in the effects of different planning mechanisms on the scalability and reward. 
To focus on the effects of planning, the three self-healing approaches that we compare to each other share the identical monitoring, analysis, and execution phases of the MAPE-K feedback loop, and they use the same architectural runtime model and utility function (knowledge). 
Moreover, the experiments are driven by deterministic failure traces, constructed from failure profile models, that enable replicating a simulation for the different approaches and over multiple runs. This allows us to fairly compare the approaches and to take variations of the measured execution time for the planning into account. 
Thus, we used various failure profile models and traces to investigate different effects of planning on scalability and reward. For instance, to investigate scalability, we focused on scaling the FGS of injected failures through the traces and the size of the \mRUBiS architecture through the \mRUBiS simulator.

Moreover, we conducted multiple specific experiments that aim for either an analytical purpose, scalability, or reward, that consider either single or multiple MAPE-K runs, and that satisfy either all or only a subset of the assumptions so that the results and their interpretation are always focused to concrete questions without confounding different effects and aspects of our overall evaluation. Finally, we followed the benchmark guidelines proposed by \citeN{Sestoft+2013} in all experiments to obtain trustworthy measurements and results.

\subsubsection{External Validity.}

Threats to external validity may restrict the generalization of our evaluation results outside the scope of our experiments. Such threats are the use of only one system under adaptation, the specific failure profile models, the specific utility function, and the three self-healing approaches.
To mitigate these threats, we use \mRUBiS as the system under adaptation that allows the injection of generic architectural failures and the repair of these failures by generic architectural adaptation rules. We can consider \mRUBiS as a generic and representative exemplar for architectural self-healing. 
Besides the synthetic, we especially use realistic failure profile models that originate from real-world systems~\cite{Gallet2010} and thus characterize how software system failures occur and propagate in practice. We even extended these realistic models to cover edge cases such as single, isolated failures or big bursts with up to about $450$ failures. Thus, we have some confidence that our evaluation results hold for real-world failure behavior.
The threat of using a specific utility function is negligible in our opinion since the same function is used for all three self-healing approaches. Thus, the utility function does not cause any effect that differs between the approaches, which could otherwise influence the results, so that we expect similar results with any other utility function.
We compared our u-driven approach to two other self-healing approaches in our experiments so that the relative results cannot be generalized to other approaches. We selected these two approaches since they cover edge cases: the static approach scales very well but often achieves non-optimal rewards, whereas the solver approach typically achieves optimal rewards but does not scale due to the costly solving of the optimization problem. Still, considering these edge cases, we can conclude that our u-driven approach is both scalable and optimal in creating plans for self-healing.

Finally, a major threat to external validity is the use of simulation instead of an actual system. However, to the best of our knowledge, simulation is the only means to evaluate the performance of self-healing systems in research.\footnote{Examples of approaches  that use simulation to evaluate self-healing systems are: \citeN{Griffith:2009:MES:1555228.1555245,Ippoliti:2012:SSA:2371536.2371551,Schmitt:2011:RSP:1998582.1998620,Ehlers:2011:SSS:1998582.1998628,Garlan&Schmerl2002,Salehie:2006:CMS:1137677.1137701,Neti:2007:QCA:1270237.1270319,5069080,5069070,1571601,Carzaniga:2008:SMA:1370018.1370023,6224391,Casanova:2013:DAR:2487336.2487354,Angelopoulos:2014:DMF:2593929.2593936,Anaya:2014:PAA:2593929.2593941,Haupt:2012:TMS:2666795.2666817,Hassan:2015:MNS:2821357.2821383,Magalhaes:2015:SWS:2744297.2700325,6063492,DiMarco:2013:SSC:2487336.2487358,Perino:2013:FSS:2486788.2487016}.}

In this context, we can categorize the state of the art in self-healing systems as work that relies on simulation and does not use failure traces at all\footnote{Examples of such work are: \citeN{Schmitt:2011:RSP:1998582.1998620,Ehlers:2011:SSS:1998582.1998628,Salehie:2006:CMS:1137677.1137701,Neti:2007:QCA:1270237.1270319,1571601,6224391,Haupt:2012:TMS:2666795.2666817}.} or work that relies on simulation and uses failure traces although these traces are not real-world traces\footnote{Such approaches either use observed and manually adjusted failure traces (e.g., \citeN{Ippoliti:2012:SSA:2371536.2371551,Garlan&Schmerl2002,5069080}), probabilistic or simple random failure traces (e.g., \citeN{Anaya:2014:PAA:2593929.2593941,6063492,5069070}), or deterministic failure traces (e.g., \citeN{Carzaniga:2008:SMA:1370018.1370023,Casanova:2013:DAR:2487336.2487354,Angelopoulos:2014:DMF:2593929.2593936,Magalhaes:2015:SWS:2744297.2700325,Perino:2013:FSS:2486788.2487016,DiMarco:2013:SSC:2487336.2487358, Griffith:2009:MES:1555228.1555245,Hassan:2015:MNS:2821357.2821383}).}. 
Thus, as shown in our previous work~\cite{8792015}, we can at first conclude that simulation is the dominating approach in the literature to evaluate self-healing systems, which confirms the general finding for self-adaptive systems by \citeN{Weyns:2012:CSE:2666795.2666811}.
Second, our findings indicate the lack of appropriate methods to evaluate self-healing systems, as we did not find any approach with performance claims that provides a complete failure profile either as representative real-world test traces or models for occurrences of failures.
This distinguishes our work from existing work, as we use realistic failure profile models for occurrences of failures, allowing us to systematically and extensively evaluate our approach using real-world data. Thus, our comparative study is more comprehensive than state-of-the-art evaluations for self-healing systems.

\subsubsection{Construct Validity.}

The major threats to construct validity are the correctness of the simulation environment, our implementation of the self-healing approaches, our adaptation of the realistic failure profile models, and our construction of the failure traces from these models. To address these threats, we use \mRUBiS as our simulation environment, which has been accepted as an exemplar by the research community on self-adaptive software and has been extensively tested by students in the scope of four courses on self-adaptive software~\cite{2018-mRUBiS}. Moreover, the implementations of the three self-healing approaches have been tested with the \mRUBiS simulator, whereas the adapted failure profile models and the traces constructed from all failure profile models have been double-checked by two authors of the article.

\section{Related Work}\label{sec:related}

As related work of this study, we discuss how the trade-off between aiming for an optimal repair and settling for a quick and efficient adaptation is practiced by planning mechanisms of self-adaptive software. 
On the one end of the spectrum, there are optimization-based approaches using runtime reasoning. \textit{All} potential adaptation decisions are determined and then evaluated at runtime by an objective function, which encounters scalability and efficiency issues~\cite{1691383,5069076}. Employing utility functions and utility-driven decision-making schemes have been extensively investigated. \citeN{Franco+2016} addresses the runtime disruption of non-functional goals by predicting their expected values for each adaptation strategy. The quantitative prediction is based on a mathematical model translated from a model of the software architecture.
We use an analytically defined utility function to compute the impact of the adaptation rules at runtime. Although the values for the parameters of our utility function are captured and updated at runtime (in the runtime model), the function itself is defined at design time. In recent work~\cite{SGCAHG18}, we learn a prediction model for the utility values instead of defining a utility function analytically.
Besides utility, our scheme also considers the execution time of individual adaptation rules when selecting and ordering such rules. Currently, we do not distinguish the execution time and latency of rules (i.e., the time until an adaptation shows an effect in the system after its execution) as we assume immediate adaptation~(repair) effects. To explicitly consider latency similarly to \citeN{7573126}, the latency of each repair rule needs to be estimated and then added to the execution time of the rule.

MOSAICO targets the trade-off between quality and computation cost by discretizing the system and environment states, and offline synthesis of adaptation plans for the different discretization points~\cite{Camara2018}. The synthesis is realized by probabilistic model checking. Similar to our~work, a utility profile steers the synthesis of plans. However, our adaptation scheme is event based rather than state based. This enables an incremental utility calculation that is based \emph{only} on the events (changes) and not on the whole state (architecture), which achieves low computation costs.

MADAM/MUSIC is an adaptive middleware for component-based applications that plans architectural adaptation by exploiting quality properties of alternative implementations of components~\cite{RouvoyEtAl09,1128711}. A QoS-aware runtime model provides the knowledge for planning adaptation that aims for maximizing the utility of the application's architecture. Using properties and property predictor functions of alternative components, each reconfiguration is planned and then evaluated for the current execution context by a utility function. The reconfiguration with the highest utility is selected for execution. 
\citeN{5069076} use reinforcement learning for online planning. 
FUSION~\cite{Esfahani+2013} also uses learning to solve the optimization problem of finding the optimal set of features that maximizes the utility. 
Such learning-based approaches suffer from a slow learning curve and achieve suboptimal utility or QoS during the time when the learning has not converged yet. 
Furthermore, probabilistic model checking has been used to solve complex optimization problems at runtime~\cite{Camara:2015:OPA:2695664.2695680, Camara:2016:AIE:2955837.2956045,Sykes:2007:PAC:1292316.1292318}. The time complexity of model checking typically results in solutions that do not scale for large configuration spaces and that cannot be applied in systems requiring instantaneous adaptation decisions.
To improve runtime efficiency, techniques such as caching, pre-computation, and near-optimality have been applied~\cite{Gerasimou:2014:ERQ:2593929.2593932}, and computations are performed as much as possible offline to reduce the planning efforts online~\cite{7573126}.
Moreover, \citeN{Moreno:2017:DCS:3105503.3105519} propose a method for combinatorial optimization based on cross-entropy and an any-time algorithm with random sampling from the solution space. 
Such solutions considerably reduce the computation time; however, they are not guaranteed to find an optimal adaptation plan. 
Summing up, utility-driven approaches pursue a search-based optimization in the solution space, which typically do not scale well for complex systems with large configuration spaces. Such approaches can manage to find the optimal configuration, but there is no guarantee to reach the result within a reasonable time. Executing an optimization algorithm for each adaptation decision at runtime causes a large overhead degrading the performance. 

In our earlier work~\cite{TG04_ag}, we suggested reducing the search space to speed up adaptation and avoid long delays.
In contrast, the self-healing scheme proposed in this article computes the utility for each possible adaptation option \textit{incrementally} at runtime taking into account the actual issues and their contexts (i.e.,~runtime knowledge influencing the utility). Due to the incremental computation, our scheme is scalable without having to reduce the search space for self-healing while achieving optimal repair plans in terms of achieved utility and reward. 

Although utility-driven, optimization-based approaches are the one end of the spectrum for decision making in self-adaptive software, whereas the other hand refers to pure rule-based approaches~\cite{1691383}. 
Rule-based approaches are recognized to be efficient and stable in predictable domains and support early validation~\cite{1537890}. They provide a quick recovery from a goal violation. However, they often result in sub-optimal adaptation decisions because they do not handle situations that have not been foreseen at design time~\cite{OwenCheng2008}. In this context, RAINBOW applies utility theory in combination with a stochastic model of the possible outcomes of the reasoning process~\cite{2012ChengStitch}. 
Whereas in our approach the effect of all possible adaptation rules on the utility is dynamically computed at runtime to select the best rules for execution, RAINBOW considers the success rate of rules in the past to rank them and to eventually make a decision. 
For that, RAINBOW uses pre-defined adaptation strategies based on the current state in the configuration space---for instance, for each observed configuration, there is a specific adaptation plan assigned at design time. In our approach, additionally to the dynamic properties of the rules in terms of utility impact and execution costs, the actual failures and their contexts (i.e., the affected components) are considered to make an adaptation decision. 

We distinguish our approach from rule-based approaches because we add runtime properties to the adaptation rules and make them event based (event-condition-action rules) rather than state based, meaning that the adaptation rules capture the change events instead of being pre-assigned to certain system configurations. 
In our approach, each adaptation rule has an initial condition, enabled by a change (event) in the system. This condition needs to be satisfied so that the rule becomes applicable. However, the condition does not provide enough information to drive the adaptation process. Thus, as opposed to any static rule-based approach, we dynamically assign applicable rules to issues based on the runtime estimations of their execution costs and their impacts on the utility.

A hybrid planning approach is proposed by~\citeN{7774394} by combining a fast, deterministic planner with a slow but optimal planner. The rationale is that the fast planner generates immediate responses while a Markov decision process planner runs in the background to look for optimal plans. 
The need for adaptation is detected by periodical evaluation of the system utility. For both planners, the adaptation plans are generated based on the current state by choosing among a set of pre-defined tactics that are applicable in the current state.
Although the idea of making a trade-off between timeliness and optimality is explored by both \citeN{7774394} and our work, we distinguish our work based on the following grounds.
Our scheme is event driven by reacting to change events so that it avoids a continuous utility evaluation to detect adaptation needs. A complete evaluation of the system utility can be costly if the architecture is large. 
Our proposed scheme seeks timeliness by using techniques for the incremental detection of adaptation issues and pattern-based computation of utility changes driven by change events.
We compute the impact of different adaptation plans (rules) at runtime regarding the change events and plan the adaptation accordingly. Due to these characteristics of our scheme, we can guarantee optimal decisions and scalability at runtime independent of the size of the architecture.

The second ground based on which we distinguish our work from~\citeN{7774394} is the treatment of time and guarantees of optimality in time-restricted adaptation loops. Using estimates for inter-arrival rates of future adaptation issues~(requests in~\cite{7774394}) defines a timing threshold to switch between the planners. In our scheme, the estimates of rule execution times~(\elem{costs}) and the planning time can be used to enforce the notion of time in terms of allowing only execution of $k$ adaptation actions. Short planning times in the approach proposed by~\citeN{7774394} can prevent the switching to the optimal planner, and hence the system operates with non-optimal utility. However, in our scheme, even with short planning times, we manage to address the failures in an optimal manner~(i.e.,~selecting optimal rules and preserving optimal ordering). The limited time will only affect the number of the failures that we resolve in one feedback run but not the optimality of the solution.

Finally, our approach is distinguished from existing work because it is scalable and optimal. Scalability is achieved by using rules and optimal decisions are guaranteed by using a utility function to drive the adaptation. Unlike other optimization-based approaches, the utility-driven process in our approach scales as the utility is computed incrementally.

\section{Conclusion and Future Work}\label{sec:conclusion}

Achieving optimal adaptation decisions online within a reasonable time is an important challenge addressed by this work. We presented a novel adaptation scheme for architectural self-healing that combines concepts of utility-driven and rule-based approaches to achieve the benefits of each of them---that is, achieving optimal adaptation decisions and being scalable. As a consequence, our combined adaptation scheme improves the scalability and reward in self-healing.

This contribution is achieved by defining the utility function and the adaptation rules in a pattern-based way, which allows us to combine the utility and the rules and therefore to compute the impact of applying each adaptation rule on the overall utility. Based on these computations and the knowledge about the execution costs of each adaptation rule, we determine and execute at runtime the optimal sequence to apply adaptation rules for self-healing. 
To evaluate the benefits of our adaptation scheme, we conducted experiments with synthetic and realistic failure profile models using the \mRUBiS simulator, in which our scheme competes with a static (rule-based) and a solver-based (utility-driven) approach to self-healing. These experiments demonstrate that our scheme achieves a considerably improved reward compared to the static approach while only having a negligible overhead. Moreover, we demonstrate that our scheme drastically reduces the computation efforts for planning self-healing compared to the solver approach when both perform optimal adaptation decisions. Being incremental makes our adaptation scheme more scalable as it faces less overhead, which becomes especially relevant for large architectures or when many failures occur, for instance, in bursts.

Finally, the incremental analysis and planning scheme presented in this article complements our earlier work on incremental monitoring and execution phases with architectural runtime models~\cite{Vogel+2009,VogelNHGB10,VG10} so that we can close the feedback loop and achieve incremental self-adaptation throughout the feedback~loop.

The presented adaptation scheme has limitations that we will address in future work. 
First of all,~the limitations refer to our assumptions (cf.~Section~\ref{subsec:discussions-for-assumptions}). We want to explore whether similar or sufficiently good (not necessarily optimal) results can be achieved by relaxing these assumptions. We already started this line of research by relaxing assumption (A2) with probabilistic adaptation rules in Section~\ref{subsec:violation-of-assumption-A2}, and we plan to continue this line by relaxing the other assumptions.
Moreover,~our~scheme uses a utility function that has been manually and analytically defined at design time. However, constructing a utility function in such a way is challenging due to various~sources of uncertainty, such as non-linearities, complex dynamic architectures, and black-box models. To address this issue, we train prediction models for the utility of systems to replace the manually and analytically defined utility functions~\cite{SGCAHG18}, and we want to study how such prediction models can be integrated into our scheme to learn and evolve utility functions online for dynamic architectures.
Finally, we want to investigate the concurrent execution~of~adaptation rules and to broaden the spectrum of self-adaptive systems to which our scheme could be applied by studying other systems than \mRUBiS and other self-* properties than self-healing.

\begin{acks}
The authors would like to thank Christian M. Adriano for assistance with the statistical bootstrapping technique and comments that improved the article.
\end{acks}

\bibliographystyle{ACM-Reference-Format}
\bibliography{references}                        

\end{document}